\documentclass[12pt,3p,times]{elsarticle}

\usepackage{amsmath,amsfonts,amssymb}
\usepackage{amsthm}
\usepackage{bm}
\usepackage{graphicx,epstopdf}
\usepackage[position=top,labelfont=normalfont,textfont=normalfont,singlelinecheck=off,justification=raggedright]{subfig}
\usepackage{floatrow}
\usepackage[dvipsnames]{xcolor}
\usepackage{setspace}
\usepackage{enumerate,etaremune}
\usepackage{booktabs}
\usepackage{tabularx,multirow}
\usepackage{framed}
\usepackage{hhline}
\usepackage{url}
\usepackage[unicode,bookmarks=false]{hyperref}
\hypersetup{
  colorlinks,
  citecolor=Blue,linkcolor=Blue,urlcolor=Blue
}

\usepackage[textsize=footnotesize,linecolor=red,backgroundcolor=red!25,bordercolor=red]
  {todonotes}

\usepackage{tikz}
\usetikzlibrary{shapes.geometric}
\usetikzlibrary{shapes,arrows}
\usetikzlibrary{plotmarks}

\usetikzlibrary{calc}

\usepackage{physics}

\usepackage{algorithm}
\usepackage{algorithmicx,algpseudocode}
\usepackage{cases}

\newcommand{\eg}{{\it e.g.}}
\newcommand{\ie}{{\it i.e.}}

\newcommand{\etal}{{\it et al.}}
\newcommand{\tensor}[1]{\bm{#1}}
\newcommand{\stress}{\sigma}

\newcommand{\pd}{\partial}

\newcommand{\el}{\mathrm{e}}
\newcommand{\pl}{\mathrm{p}}
\newcommand{\trial}{\mathrm{tr}}
\newcommand{\rn}[1]{\uppercase\expandafter{\romannumeral #1\relax}}
\newcommand{\cn}{\mathrm{N}}

\newcommand{\jrc}{\mathrm{JRC}}
\newcommand{\jcs}{\mathrm{JCS}}

\DeclareMathOperator{\dyadic}{\otimes}
\newsavebox{\dotbox}

\theoremstyle{remark}
\newtheorem{remark}{Remark}

\newcommand{\revised}[1]{{\color{black} #1}}

\newcolumntype{L}[1]{>{\raggedright\let\newline\\arraybackslash\hspace{0pt}}m{#1}}
\newcolumntype{C}[1]{>{\centering\let\newline\\arraybackslash\hspace{0pt}}m{#1}}
\newcolumntype{R}[1]{>{\raggedleft\let\newline\\arraybackslash\hspace{0pt}}m{#1}}

\allowdisplaybreaks

\biboptions{sort&compress,square,comma,numbers}
\AtBeginDocument{\hypersetup{citecolor=MidnightBlue,linkcolor=MidnightBlue,urlcolor=MidnightBlue}}

\usepackage{etoolbox}
\makeatletter
\patchcmd{\ps@pprintTitle}
{Preprint submitted to}
{~}
{}{}
\makeatother

\begin{document}

\begin{frontmatter}

\title{Extended Barton--Bandis model for rock joints under cyclic loading: Formulation and implicit algorithm}

\author[LLNL]{Fan Fei}
\author[KAIST]{Jinhyun Choo\corref{corr}}
\cortext[corr]{Corresponding Author}
\ead{jinhyun.choo@kaist.ac.kr}

\address[LLNL]{Atmospheric, Earth, and Energy Division, Lawrence Livermore National Laboratory, United States}
\address[KAIST]{Department of Civil and Environmental Engineering, KAIST, South Korea}

\journal{~}

\begin{abstract}
In this paper, the Barton--Bandis model for rock joints is extended to cyclic loading conditions, without any new material parameter.
Also developed herein is an algorithm for implicit numerical solution of the extended Barton--Bandis model, which can also be used for the original Barton--Bandis model for which an implicit algorithm has been unavailable.
To this end, we first cast the Barton--Bandis model into an incremental elasto-plastic framework, deriving an expression for the elastic shear stiffness being consistent with the original model formulation.
We then extend the model formulation to cyclic loading conditions, incorporating the dependence of shear stress and dilation on the joint position and the shearing direction.
The extension is achieved by introducing a few state-dependent variables which can be calculated with the existing material parameters.
For robust and accurate utilization of the model, we also develop an implicit algorithm based on return mapping, which is unconditionally stable and guarantees satisfaction of the strength criterion.
We verify that the proposed model formulation and algorithm produces the same results as the original Barton--Bandis model under monotonic shearing conditions.
We then validate the extended Barton--Bandis model against experimental data on natural rock joints under cycling loading conditions.
The present work thus enables the Barton--Bandis model, which has been exceptionally popular in research and practice, to be applicable to a wider range of problems in rock mechanics and rock engineering.
\end{abstract}

 
\end{frontmatter}


\section{Introduction}
\label{sec:intro}

The Barton--Bandis model~\cite{barton1977shear,bandis1983fundamentals,barton1985strength} is a physically-motivated empirical constitutive model for rock joints, which has been extensively used for a wide range of rock mechanics problems in research and practice.
The unparalleled popularity of the model can be attributed to the combination of the following two features:
(i) it can well reproduce the shearing and dilation responses of rock joints and their dependence on normal stress and joint roughness, 
and (ii) its material parameters, such as the residual friction angle, the joint roughness coefficient (JRC), and the joint compression strength (JCS), are all physically meaningful and commonly characterized in practice.

When it comes to rock joints under cyclic loading, however, the Barton--Bandis model faces some limitations.
Since the model was originally formulated for monotonic loading, it does not incorporate an array of behavior characteristics that depend on the joint position and/or the shearing direction (see \eg~\cite{hutson1990joint,huang1993investigation,homand2001friction,lee2001influence,meng2018characteristics} for experimental evidence). 
For this reason, when Barton~\cite{barton1982modelling} developed the concept of mobilized joint roughness for describing pre- and post-peak monotonic shearing behavior and its size dependence, he also presented an attempt to extend the original model to cyclic loading based on the mobilized joint roughness coefficient.
However, this attempt was not concerned with contractive volume change behavior during cyclic loading and was validated against a single dataset only. 
Later, Asadollahi and Tonon~\cite{asadollahi2011degradation} proposed a modified Barton--Bandis model based on the results of eighteen cyclic shear box tests. 
While the modified model was thoroughly validated, it is rather complex and not the same as the Barton--Bandis model under monotonic shearing conditions.
Thus it would require significant effort to implement and use the modified model for practical problems.

From a numerical perspective, the lack of an implicit algorithm for the Barton--Bandis model has been a critical hurdle for applying the model to every kind of problem that it can benefit.
Due to its empirical nature, the Barton--Bandis model is not formulated in an incremental (rate) form.
Therefore, although the Barton--Bandis model has been implemented in explicit numerical methods (\eg~\cite{choi2004stability,lei2016implementation,ma2017implementation}), it has not been compatible with implicit methods such as finite elements for discontinuities (\eg~\cite{liu2008contact,fei2020phasea,fei2021phase,choo2021barrier}). 
This limitation has hampered the proper use of the Barton--Bandis model for many important problems that involve a relatively long period of time and/or complex coupling with environmental processes such as fluid flow and heat transfer.
It also restricts the applicability of the Barton--Bandis model under general cyclic loading conditions, because cyclic shearing is often triggered by changes in fluid pressure and/or temperature over a long period of time (\eg~\cite{bettinelli2008seasonal,zhou2015damage,preisig2016hydromechanical}).
It is also noted that the use of an implicit integration algorithm can significantly improve the robustness and accuracy of numerical simulation of rock masses; see, \eg~Hashimoto \etal~\cite{hashimoto2021improvement} in the context of discontinuous deformation analysis.
Therefore, an implicit algorithm for the Barton--Bandis model, similar to those developed for a variety of other constitutive models (\eg~\cite{borja1990cam,borja2003numerical,white2014anisotropic,borja2016cam,choo2018mohr}), is highly desired.

Motivated by the above-described limitations of the Barton--Bandis model, the work has two objectives: (i) to extend the model formulation to cyclic loading conditions with minimal ingredients, and (ii) to develop an implicit algorithm that can be used for both the original and extended Barton--Bandis model.
To achieve the first objective, we introduce a few state-dependent variables---but not a new material parameter---capturing the position- and direction-dependence of shear stress and dilation in rough rock joints.
To achieve the second, we cast the Barton--Bandis model into an incremental elasto-plastic framework and develop an implicit return mapping algorithm applicable to every type of shearing stage.
The proposed formulation and algorithm is verified to produce virtually the same results as the original Barton--Bandis model under monotonic shearing conditions.
The extended Barton--Bandis model is then validated to be in good agreement with experimental data on the shear stress and dilation behavior of natural rock joints under cyclic loading conditions.

\section{Incremental elasto-plastic formulation of the Barton--Bandis model}
\label{sec:reformulation}

In this section, we cast the Barton--Bandis model into an incremental elasto-plastic framework for rock joints.
The purpose of this elasto-plastic formulation is to make the Barton--Bandis model compatible with a robust implicit integration algorithm, which will be developed later in this paper.

\subsection{General constitutive law and kinematics}
The general incremental form of a joint constitutive law can be written as
\begin{equation}
  \dot{\bm{t}} = \mathbb{C}\cdot\dot{\bm{u}},
\end{equation}
where $\bm{t}$ is the traction vector, $\bm{u}$ is the displacement vector, and $\mathbb{C}$ is the tangent stiffness of the joint.

As standard, we first decompose the joint displacement vector into its normal and shear directions as
\begin{align}
    \tensor{u} = u_{\cn} \tensor{n} + \delta \tensor{m}
  \label{eq:u-decompose-direction}
\end{align}

where $u_{\cn}$ and $\delta$ are the magnitudes, and $\tensor{n}$ and $\tensor{m}$ are the unit vectors, of the joint normal and shear displacements, respectively. 
Applying the same directional decomposition to the traction vector gives  
\begin{align}
  \tensor{t} = \stress_{\cn} \tensor{n} + \tau \tensor{m} , 
  \label{eq:t-decompose-direction}
\end{align}
where $\stress_{\cn}$ and $\tau$ are the joint normal and shear stresses, respectively. 
In this work, we consider compressive stress positive following the standard sign convention in rock mechanics.

Elasto-plastic modeling of rock joints begins by postulating that the joint displacement can be additively decomposed into elastic and inelastic (plastic) parts as
\begin{align}
  \tensor{u} = \tensor{u}^{\el} + \tensor{u}^{\pl} . 
  \label{eq:u-decompose-elastoplastic}
\end{align}
where superscripts $(\cdot)^{\el}$ and $(\cdot)^{\pl}$ denote the elastic and inelastic parts, respectively. 
Accordingly, the joint normal and shear displacements are decomposed into their elastic and inelastic parts as
\begin{align}
    u_{\cn} &= u^{\el}_{\cn} + u^{\pl}_{\cn} \\ 
    \delta &= \delta^{\el} + \delta^{\pl}  
\end{align} 
The elastic part corresponds to the recoverable joint displacement associated with the interlocking behavior of rough joints. 
The inelastic part represents the unrecoverable joint displacement which involves the mobilization of joint roughness. 

In what follows, we introduce constitutive relationships to the elastic and inelastic joint deformations based on the Barton--Bandis model.
We first consider the inelastic deformation as it can be fully described by Barton's strength criterion.

\subsection{Inelastic deformation}
Let us first recapitulate a general plasticity framework for rock joints.
The shear strength of a joint is described by a yield function $F$, given by
\begin{align}
  F(\tensor{t}) = \tau - \stress_{\cn} \tan \phi \leq 0 , 
  \label{eq:yield-function}
\end{align}
where $\phi$ is the friction angle, which may vary with the normal stress and the shear displacement.
As a non-associative flow rule is supported by experimental evidence, a potential function $G\neq F$ is introduced, of which the general form is given by~\cite{son2004elasto}
\begin{align}
  G(\tensor{t}) = \tau - \int \tan \psi \: \dd \stress_{\cn}, 
  \label{eq:potential-function}
\end{align}
where $\psi$ is the dilation angle. 
The flow rule gives the rate of the inelastic displacement as
\begin{align}
  \dot{\tensor{u}}^{\pl} = \dot{\lambda} \dfrac{\pd G}{\pd \tensor{t}} = \dot{\lambda} \left(\tensor{m} - \tensor{n} \tan \psi \right) , 
  \label{eq:flow-rule}
\end{align}
where $\lambda$ denotes the plastic multiplier. 

We now specialize the framework to the Barton--Bandis model. 
Recall that the Barton--Bandis strength criterion is given by
\begin{align}
  \tau = \stress_{\cn} \tan \left[\phi_r + \jrc_\mathrm{m} \log \left(\dfrac{\jcs}{\stress_{\cn}} \right) \right] , 
  \label{eq:shear-strength-barton}
\end{align}
where $\phi_r$ is the residual friction angle, $\jrc_\mathrm{m}$ is the mobilized joint roughness coefficient, and $\jcs$ is the joint wall compressive strength~\cite{barton1977shear}.
The residual friction angle can be empirically related to the basic friction angle, $\phi_\mathrm{b}$, which can be obtained from tilt tests in the laboratory. 
For example, Barton and Choubey~\cite{barton1977shear} proposed the following empirical relationship between $\phi_r$ and $\phi_b$
\begin{equation}
  \phi_{r} = (\phi_{b} - 20) + 20(r_{1}/r_{2}) \quad \text{(in degrees),}
\end{equation}
where $r_1$ is the Schmidt rebound number on wet and weathered joint surfaces, and $r_2$ is the Schmidt rebound number on dry and unweathered surfaces.
Notably, $\phi_{r}$ is often assumed to be equal to $\phi_{b}$ if there is no weathering.
To make the yield function~\eqref{eq:yield-function} equivalent to the Barton--Bandis strength criterion~\eqref{eq:shear-strength-barton}, we set the total friction angle as
\begin{align}
  \phi = \phi_r + \jrc_\mathrm{m} \log \left(\dfrac{\jcs}{\stress_{\cn}} \right) \quad \text{(in degrees).}
  \label{eq:friction-angle} 
\end{align} 
The dilation angle of the Barton--Bandis model is given by~\cite{olsson2001improved}, 
\begin{align}
  \psi = \dfrac{1}{M} \jrc_\mathrm{m} \log\left(\dfrac{\jcs}{\stress_{\cn}} \right) \quad \text{(in degrees)}, 
  \label{eq:dilation-angle}
\end{align}
where $M$ is the damage coefficient accounting for asperity degradation due to shearing. 
The damage coefficient can be estimated by an empirical equation proposed by Barton and Choubey~\cite{barton1977shear}, which is written as 
\begin{align}
  M = 0.7 + \dfrac{\jrc_p}{12 \log\left[\jcs/\stress_{\cn} \right]},
  \label{eq:damage-coeff}
\end{align}
where $\jrc_p$ is the peak joint roughness coefficient~\cite{barton1982modelling,barton1985strength}.
Note that the damage coefficient increases with the magnitude of the normal stress. 
This stress dependence of $M$ allows one to capture that a joint experiences more asperity damage under a larger confining pressure, see \eg~\cite{barton1971relationship,lee2001influence,meng2018characteristics} for experimental evidence.
We also note that both $\jrc_{p}$ and $\jcs$ are subject to size effects~\cite{barton1980some,barton1981some,bandis1981experimental}, which can be estimated by scaling laws proposed by Barton and Bandis~\cite{barton1982effects}
\begin{align}
  \jrc_{p} &= \jrc_{p0} \left(\dfrac{l_{j}}{l_0} \right)^{-0.02 \jrc_{p0}} , \\ 
  \jcs &= \jcs_{0} \left( \dfrac{l_{j}}{l_0} \right)^{-0.03 \jrc_{p0}} , 
\end{align}
where $\jrc_{p0}$ and $\jcs_{0}$ refer to the values of $\jrc_{p}$ and $\jcs$, respectively, measured from a joint of length $l_0$, and $l_{j}$ is the length of the joint of interest.

Calculation of the friction angle~\eqref{eq:friction-angle} and the dilation angle~\eqref{eq:dilation-angle} requires one to evaluate the mobilized joint roughness coefficient, $\jrc_\mathrm{m}$, during the course of shearing. 
The original model of Barton and coworkers~\cite{barton1982modelling,barton1985strength} provides values of the normalized $\jrc_\mathrm{m}$ as a function of the normalized shear displacement, as presented in Table~\ref{tab:jrc-barton}.
In this table, $\delta_p$ is the shear displacement at which $\jrc_\mathrm{m}$ reaches its peak value, $\jrc_p$, which can be estimated as~\cite{barton1982effects,barton1985strength}
\begin{align}
  \delta_p = \dfrac{l_{j}}{500} \left(\dfrac{\jrc_p}{l_{j}} \right)^{0.33} . \label{eq:v-p}
\end{align}  
Also, parameter $i$ is defined as 
\begin{align}
  i = \jrc_p \log\left(\dfrac{\jcs}{\stress_{\cn}} \right) , 
\end{align}  
such that Eq.~\eqref{eq:shear-strength-barton} gives zero shear stress when there is no shear displacement. 
\begin{table}[h!]
  \centering
  \begin{tabular}{l|ccccccccc}
  \toprule
  $\delta/\delta_p$ & 0.0 & 0.3 & 0.6 & 1.0 & 2.0 & 4.0 & 10.0 & 25.0 & 100.0  \\
  \midrule
  $\jrc_\mathrm{m}/\jrc_p$ & $-\phi_r/i$ & 0.0 & 0.75 & 1.0 & 0.85 & 0.70 & 0.50 & 0.40 & 0.0 \\ 
  \bottomrule
  \end{tabular}
  \caption{Normalized $\jrc_\mathrm{m}$ values in the Barton--Bandis model~\cite{barton1982modelling,barton1985strength}.}
  \label{tab:jrc-barton}
\end{table}

In the literature, $\jrc_\mathrm{m}$ has usually been calculated by interpolating the values in Table~\ref{tab:jrc-barton} in a piecewise linear manner.  
However, this approach is not optimal from a numerical perspective, because the derivative of $\jrc_\mathrm{m}$ is discontinuous at the shear displacements designated in the table.
Such discontinuities in the derivatives make it tricky to evaluate the flow rule~\eqref{eq:flow-rule} and may be detrimental to the convergence behavior in an implicit numerical method.

Therefore, here we adopt a smoothed relation between $\jrc_\mathrm{m}$ and $\delta$, which has recently been proposed by Prassetyo~\etal~\cite{prassetyo2017nonlinear}.
The smoothed relation can be written as
\begin{align}
  \jrc_\mathrm{m} = 
  \begin{cases}
    \left[\dfrac{7(1 + \phi_r/i)\delta}{3 \delta_{p} - (3 - 7 \phi_r/i)\delta} - 1\right] \dfrac{\phi_r}{i} \jrc_{p} & \text{if}\; 0 \leq \delta < \delta_{p}  , \vspace{2em} \\ 
    \left[-0.217 \ln \left(\dfrac{\delta}{\delta_{p}} \right) +1 \right] \jrc_{p} & \text{if}\; \delta_{p} \leq \delta .
  \end{cases}
  \label{eq:smooth-jrc}
\end{align}
Here, $\jrc_\mathrm{m}$ is a hyperbolic function of $\delta$ when $0 \leq \delta < \delta_{p}$, and it is a logarithmic function when $\delta_{p} \leq \delta$. 

\subsection{Elastic deformation}
In the Barton--Bandis model, the elastic response in the joint normal direction is described by the hyperbolic equation proposed by Bandis~\etal~\cite{bandis1983fundamentals}, given by
\begin{align}
  \stress_{\cn} = \dfrac{\kappa u^{\el}_{\cn}}{1 - {u^{\el}_{\cn}}/u^{\el}_{\max}}, 
  \label{eq:normal-stress}
\end{align} 
where $\kappa$ is the initial normal stiffness, and $u^{\el}_{\max}$ is the maximum joint closure.  
Both $\kappa$ and $u^{\el}_{\max}$ can be evaluated using the following empirical equations~\cite{bandis1983fundamentals}
\begin{align}
  \kappa &= -7.15 + 1.75 \jrc_p + 0.02 \dfrac{\jcs}{a_{j}} , \\ 
  u^{\el}_{\max} &= 0.296 + 0.0056\jrc_p + 2.241 \left (\dfrac{\jcs}{a_{j}} \right)^{-0.245} .
\end{align}
Here, $a_{j}$ is the initial (unstressed) joint aperture, which can be approximated as~\cite{barton1982modelling}
\begin{align}
  a_{j} = \dfrac{\jrc_p}{5} \left( 0.2 \dfrac{\stress_{c}}{\jcs} - 0.1 \right),
  \label{eq:init-aperture}
\end{align}  
with $\stress_{c}$ denoting the uniaxial compressive strength of the rock. 
If there is no weathering, $\stress_{c}$ is assumed to be equal to $\jcs$~\cite{barton1977shear}, and Eq.~\eqref{eq:init-aperture} simplifies to
\begin{align}
  a_{j} = \dfrac{\jrc_p}{50} . 
\end{align}

Unlike the joint elasticity in the normal direction, the elastic joint response in the shear direction is not explicitly described in the Barton--Bandis model. 
However, an explicit description of the elastic shear stiffness is required in an elasto-plastic framework.
Therefore, in the following, we derive an expression for the elastic shear stiffness being consistent with the Barton--Bandis model, without introducing any new parameter.
The feasibility of the proposed shear stiffness will be validated later through a comparison with experimental data on cyclic shear loading.

\begin{figure}[h!]
  \centering
  \includegraphics[height=0.48\textwidth]{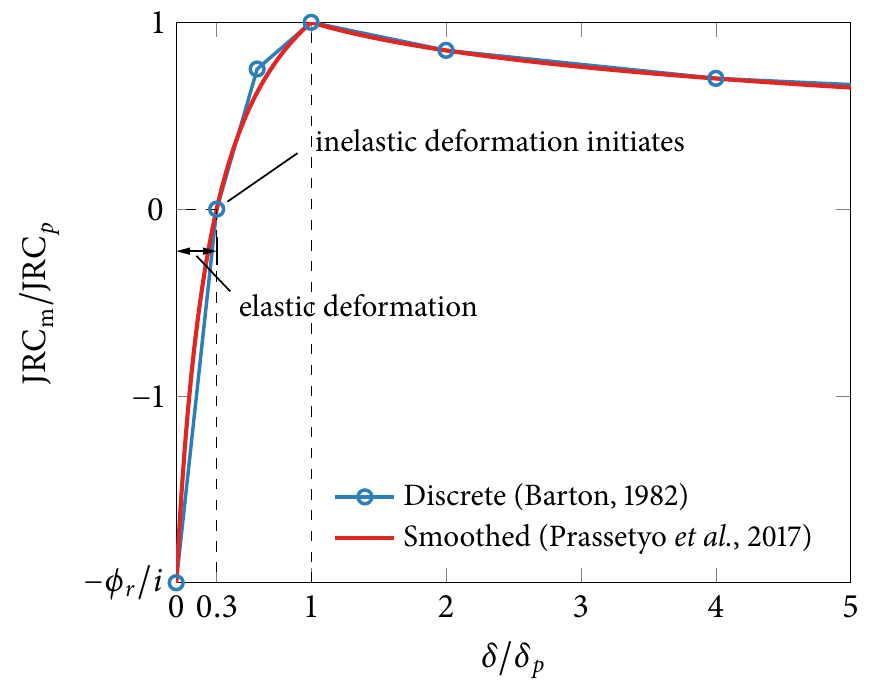}
  \caption{Elastic and inelastic regimes interpreted for the elasto-plastic formulation of the Barton--Bandis model, along with the smoothed $\jrc_\mathrm{m}$ curve proposed by Prassetyo \etal~\cite{prassetyo2017nonlinear}, Eq.~\eqref{eq:smooth-jrc}, and the discrete $\jrc_\mathrm{m}$ values in Barton~\cite{barton1982modelling}, Table~\ref{tab:jrc-barton}.}
  \label{fig:jrc-smooth}
\end{figure}

The joint elasticity in the shear direction is given by
\begin{align}
  \tau = \mu  {\delta^{\el}} , 
  \label{eq:shear-stress}
\end{align}
where $\mu$ is the \emph{elastic} shear stiffness.
To derive an expression for $\mu$ consistent with the Barton--Bandis model, we should first delineate the elastic shear regime in the Barton--Bandis model. 
As can be seen from Fig.~\ref{fig:jrc-smooth}, $\jrc_\mathrm{m}$ is negative when $\delta<0.3 \delta_p$ and $\jrc_\mathrm{m} = - \phi_r/i$ at $\delta=0$.
These negative $\jrc_\mathrm{m}$ values are intended to produce zero-dilation shearing, which takes place when the joint surfaces are being interlocked, until $\delta$ reaches $0.3\delta_p$.
Therefore, it is consistent with the Barton--Bandis model to postulate that joint shear deformation is elastic until $\delta$ reaches $0.3\delta_p$. 
We then derive the elastic shear stiffness such that the shear stresses calculated by Eqs.~\eqref{eq:shear-strength-barton} and~\eqref{eq:shear-stress} are identical when inelastic joint deformation emerges at $\delta=0.3\delta_{p}$. As $\jrc_\mathrm{m} = 0$ at this point, we get
\begin{align}
  \tau \lvert_{\delta = 0.3 \delta_{p}} =\stress_{\cn} \tan \phi_r. 
\end{align} 
Inserting this shear stress and $\delta^{\el} = 0.3\delta_{p}$ into Eq.~\eqref{eq:shear-stress} gives
\begin{align}
  \mu = \dfrac{\stress_{\cn}\tan \phi_r}{0.3 {\delta_{p}}} . 
  \label{eq:shear-modulus}
\end{align}
Note that the above expression accounts for pressure-dependent shear stiffness, which has been observed by many experimental investigations~\cite{barton1982modelling,meng2018characteristics,homand2001friction,barton1985strength}. 

\section{Extension of the Barton--Bandis model to cyclic loading}
\label{sec:extension}
In this section, we extend the Barton--Bandis model to rough rock joints under cyclic loading.
No additional material parameter is introduced, so as to retain the practical merits of the original Barton--Bandis model.

\subsection{Shear stress and dilation behavior of a joint under cyclic loading}
\begin{figure}[h!]
    \centering
    \subfloat[Shear stress]{\includegraphics[width=0.48\textwidth]{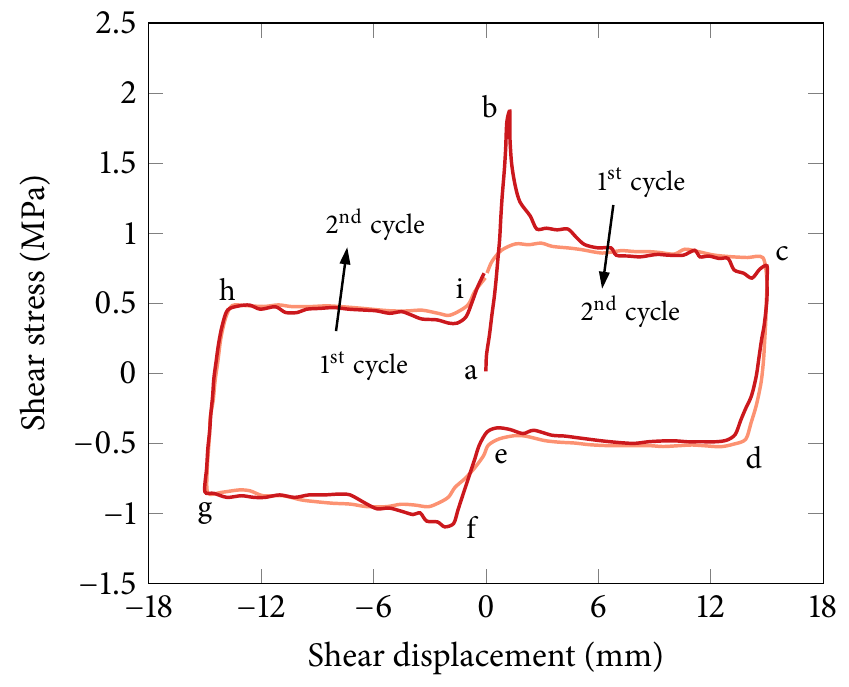}}\hspace{0.5em}
    \subfloat[Dilation]{\includegraphics[width=0.48\textwidth]{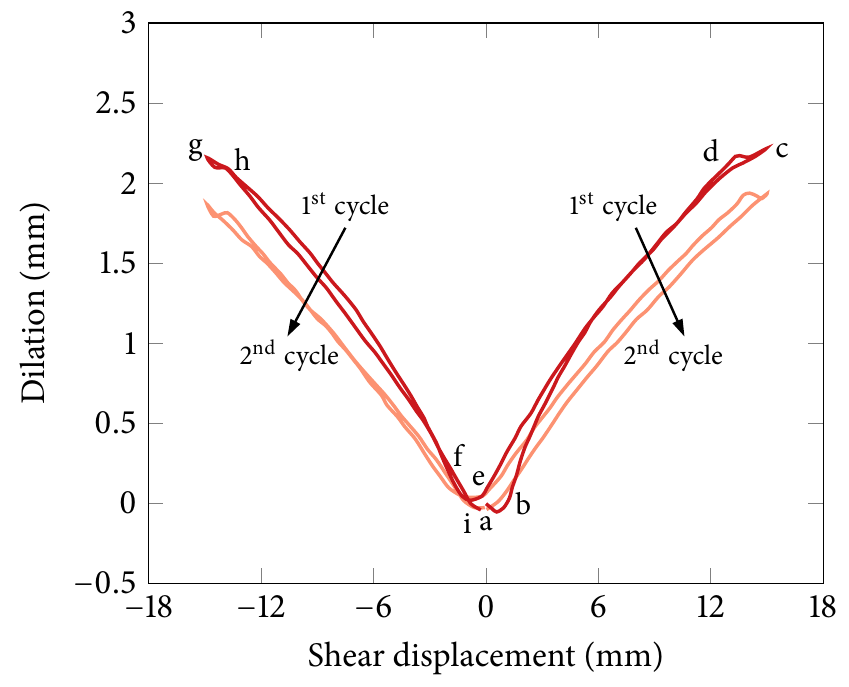}}
    \caption{Experimental results of a granite joint under cyclic loading. From Lee~\etal~\cite{lee2001influence}.}
    \label{fig:lee-cyclic-lab}
\end{figure}

Let us first review the shear stress and dilation behavior of a joint under cyclic loading. 
As an illustrative example, in Fig.~\ref{fig:lee-cyclic-lab} we show the experimental results of a granite joint obtained by Lee~\etal~\cite{lee2001influence}.
\revised{Under a normal stress of 1 MPa, the joint was sheared slowly with rates ranging 0.5--0.8 mm/s such that it is under quasi-static loading conditions.}
Note that this joint had an irregular configuration like a typical joint in natural rock masses, unlike regular-shaped joints studied in some other laboratory experiments~\cite{hutson1990joint,huang1993investigation}. 

Depending on the current joint position and the shearing direction, the load cycle in Fig.~\ref{fig:lee-cyclic-lab} can be divided into four stages.
The behavior characteristics in these four stages can be summarized as follows.
(The stage names are adopted from Lee~\etal~\cite{lee2001influence}.)

\begin{itemize}\itemsep=0pt
  \item  \textbf{Forward advance (a-b-c).} 
  In this initial loading stage, the shear stress first increases until it reaches the peak strength (a-b), and then it decreases and gradually approaches the residual shear strength (b-c). 
  Meanwhile, the joint keeps dilating after roughness is mobilized.

  \item \textbf{Forward return (c-d-e).}
  When the shearing direction is reversed, the joint first undergoes a small amount of elastic unloading (c-d) and then a reversed shearing (d-e). 
  In this stage, the joint contracts throughout and shows almost no residual volume change when it returns to the initial mated position.  
  It is noted that the shear stress in the reversed shearing phase (d-e) is almost constant and lower than the residual shear stress at the end of the forward advance stage.
  This is because dilation in the forward advance stage provides shear resistance, whereas contraction takes place in the return stage.

  \item \textbf{Backward advance (e-f-g).} After the joint passes its mated position, the shear stress increases to a new peak strength in the backward stage (e-f) followed by a softening phase (f-g). 
  This new peak strength is lower than that in the forward stage.
  Also, the joint shows a dilation behavior similar to that in the forward advance stage, but the amount of dilation is slightly lower.
  These strength and dilation responses suggest that the joint has a lower level of roughness in the backward stage, which in turn indicates that asperity damage is dependent on the direction of shear displacement.

  \item \textbf{Backward return (g-h-i).} As the shear displacement reverses in the backward stage, the joint experiences a small amount of elastic unloading (g-h) followed by a constant shear stress (h-i), similar to its response in the forward return stage. 
  Also, the joint shows contraction throughout and nearly zero volume change at its mated position.
\end{itemize}

In addition, the shearing behavior also shows some hysteresis.
Due to asperity damages during the first cycle, the amounts of both dilation and contraction are reduced in the second cycle.
This change in turn makes the shear strength lower in advance stages where dilation takes place, and make it slightly greater in return stages where contraction takes place.
The shear strength difference is particularly pronounced in the forward advance stage, because the very first peak strength in the first cycle emanated from undamaged asperities.

\subsection{State variables for the joint position and the shearing direction}
Recall that the original Barton--Bandis model is concerned with the forward advance stage only.
Therefore, to extend the Barton--Bandis model to the other three stages, we first need a quantitative approach to accounting for the current joint position (forward \textit{vs.} backward) and the shearing direction (advance \textit{vs.} return).

For this purpose, we define two types of variables (but not new material parameters).
The first is a variable that has different signs in advance and return stages.
Inspired by the formulation of White~\cite{white2014anisotropic}, we define the variable as
\begin{align}
  \alpha = {\dfrac{\dot{\delta}}{\lvert \dot{\delta} \rvert} \dfrac{\delta}{\lvert \delta \rvert}}
  \label{eq:direction-alpha}
\end{align}
By definition, $\alpha = +1$ in an advance stage and $\alpha=-1$ in a return stage.
As forward and backward stages can be distinguished by the sign of slip displacement $\delta$, the current stage can be mathematically identified as
\begin{align}
  \text{stage} =
  \begin{cases}
    \text{forward advance} & \text{if}\; {\delta} \geq 0 \;\text{and}\; \alpha = +1,\\ 
    \text{forward return} & \text{if}\; {\delta} \geq 0 \;\text{and}\; \alpha = -1,\\
    \text{backward advance} & \text{if}\; {\delta} < 0 \;\text{and}\; \alpha = +1,\\
    \text{backward return} & \text{if}\; {\delta} <0 \;\text{and}\; \alpha = -1.
  \end{cases}
  \label{eq:loading-stage}
\end{align}  
Second, we introduce two variables representing accumulated slip displacements in the forward and backward positions separately.
Specifically, we define the accumulated slip displacement in the forward stage, $\Lambda_\mathrm{f}$, and that in the backward stage, $\Lambda_\mathrm{b}$, as
\begin{align}
  \Lambda_\mathrm{f} &= 0.3{\delta_{p}} + \int_\text{inelastic} H({\delta}) H(\alpha) \lvert {\dot{\delta}} \rvert \: \dd t , 
  \label{eq:lambda-forward} \\
  \Lambda_\mathrm{b} &= 0.3{\delta_p} + \int_\text{inelastic} H({-\delta}) H(\alpha)\lvert  {\dot{\delta}} \rvert \: \dd t , 
  \label{eq:lambda-backward}
\end{align}
respectively, where $H(\cdot)$ denotes the Heaviside function, defined as 
\begin{align}
  H(x) = 
  \begin{cases} 
    1 & \text{if}\; x \geq 0, \\ 
    0 & \text{if}\; x < 0 ,
  \end{cases} 
\end{align}
and $\int_{\text{inelastic}}$ means that the integration is performed during an inelastic deformation ($F=0$).
The first Heaviside function in Eqs.~\eqref{eq:lambda-forward} and \eqref{eq:lambda-backward} is to ensure that $\Lambda_\mathrm{f}$ and $\Lambda_\mathrm{b}$ are updated only in forward and backward stages, respectively. 
The second Heaviside function is to let the accumulated slip variables remain unchanged in their corresponding return stages. 
Also, because the joint roughness evolves in the inelastic deformation regime (${\delta}\geq 0.3{\delta_{p}}$), an initial value of $0.3 {\delta_{p}}$ has been added to the accumulated slip variables. 
For notational convenience, we also define a unified variable representing the accumulated slip displacement as
\begin{align}
  \Lambda = 
  \begin{cases}
    \Lambda_\mathrm{f} & \text{if}\; {\delta} \geq 0,  \\
    \Lambda_\mathrm{b} & \text{if}\; {\delta} < 0. 
  \end{cases}
  \label{eq:lambda-unified}
\end{align}

\subsection{Extension of the shear strength to cyclic loading}
We now extend the shear strength of the Barton--Bandis model, Eq.~\eqref{eq:shear-strength-barton}, to cyclic loading.
Recall that the physical mechanism why the shear strength is different during cyclic loading is that the sense (dilative \textit{vs.} contractive) and degree of roughness mobilization depend on the joint position and shearing direction.
Also, the effect of roughness mobilization on the shear strength in the Barton--Bandis model is encapsulated in $\jrc_\mathrm{m}$. 

Therefore, we modify the sign and degree of $\jrc_\mathrm{m}$ as follows.
First, we multiply $\alpha$ defined in Eq.~\eqref{eq:direction-alpha} to $\jrc_\mathrm{m}$ such that the value of $\jrc_\mathrm{m}$ is positive in an advance stage and negative in an return stage.
In this way, $\phi > \phi_r$ in an advance stage where dilation gives rise to additional shear resistance, and $\phi < \phi_r$ in an return stage where contraction makes the shear strength lower than the residual value. 
It is noted that the same relation between $\phi$ and $\phi_r$ is common in other physically motivated models for rock joints under cyclic loading (\eg~\cite{plesha1987constitutive,jing1993study,stupkiewicz2001modeling,white2014anisotropic}).
Second, to account for that the peak strength in the backward stage is lower than that in the forward stage, we reduce the value of $\jrc_{p}$ in the backward stage.
Particularly, adopting an empirical equation in Asadollahi and Tonon~\cite{asadollahi2011degradation}, we define $(\jrc_{p})_{\tau}$---the peak value of $\jrc$ associated with the shear strength---as
\begin{align}
  (\jrc_{p})_{\tau} = 
  \begin{cases}
    \jrc_{p} & \text{if}\; {\delta} \geq 0, \\
    0.87\jrc_{p} & \text{if}\; {\delta} < 0.
  \label{eq:jrcp-strength}
  \end{cases}
\end{align}
Note that $(\jrc_{p})_{\tau}$ will only replace $\jrc_{p}$ in calculating $\jrc_\mathrm{m}$.
It will not be used for evaluating the damage coefficient~\eqref{eq:damage-coeff} and the peak shear displacement~\eqref{eq:v-p}.

In addition to the above modification, we replace ${\delta}$ in Eq.~\eqref{eq:smooth-jrc} with $\Lambda$, to account for the direction-dependence of asperity damage. This gives
\begin{align}
  \jrc_\mathrm{m} = 
  \begin{cases}
    \alpha \left[\dfrac{7(1 + \phi_r/i_{\tau})\Lambda}{3 {\delta_{p}} - (3 - 7 \phi_r/i_{\tau})\Lambda} - 1\right] \dfrac{\phi_r}{i_{\tau}} (\jrc_{p})_{\tau} & \text{if}\; 0 \leq \Lambda < {\delta_{p}}  , \vspace{2em}  \\ 
    \alpha \left[-0.217 \ln \left(\dfrac{\Lambda}{{\delta_{p}}} \right) +1 \right] (\jrc_{p})_{\tau} & \text{if}\; {\delta_{p}} \leq \Lambda ,
  \end{cases}
  \label{eq:smooth-jrc-cyclic}
\end{align}
where
\begin{equation}
  i_\tau = (\jrc_p)_{\tau} \log\left(\dfrac{\jcs}{\stress_{\cn}} \right).
  \label{eq:parameter-i-cyclic}
\end{equation}
Finally, inserting Eq.~\eqref{eq:smooth-jrc-cyclic} into Eq.~\eqref{eq:shear-strength-barton}, the shear strength is extended to cyclic loading.

It is noted that here we have used the same function form of $\jrc_\mathrm{m}$ (but with different signs determined by $\alpha$) for both forward and backward stages, whereas Asadollahi and Tonon~\cite{asadollahi2011degradation} proposed a new function for the evolution of $\jrc_\mathrm{m}$ in the backward stage.

\subsection{Extension of the dilation angle to cyclic loading}
We further extend the dilation angle of the Barton--Bandis model, which was originally proposed for the forward advance stage, to the other three stages.
Firstly, we recall that the dilation behavior in the backward advance stage is qualitatively similar to that in the forward stage, while the amount of dilation is lower. 
Therefore, the dilation angle in both the forward and backward advance stages can be expressed as the original equation~\eqref{eq:dilation-angle}, using $\jrc_\mathrm{m}$ in Eq.~\eqref{eq:smooth-jrc-cyclic} which accounts for the reduced peak mobilization in backward stages via $(\jrc_{p})_{\tau}$~\eqref{eq:jrcp-strength}.
Next, for the dilation angles in the forward and backward return stages, we assume that it is negative (contractive) and its magnitude decreases linearly to zero after elastic unloading such that there is no residual volume change in the initial mated position.
A specific expression for this dilation angle in return stages can be obtained by combining the flow rule~\eqref{eq:flow-rule} and the directional decomposition~\eqref{eq:u-decompose-direction}.  
Eventually, we get
\begin{align}
  \psi = 
  \begin{cases}
    \dfrac{1}{M} \jrc_\mathrm{m} \log\left(\dfrac{\jcs}{\stress_{\cn}} \right) & \text{if}\; \alpha = +1 , \vspace{2em} \\
    - \arctan \left( \dfrac{{u^{\pl}_{\cn}}}{ \lvert {\delta} \rvert} \right)\dfrac{180}{\pi} & \text{if}\; \alpha = -1,
  \end{cases}
  \quad \text{(in degrees).}
  \label{eq:dilation-angle-cyclic}
\end{align}
Note that $180/\pi$ is multiplied to the second equation to express the dilation angle in degrees, as in the original Barton--Bandis model.

\subsection{Comparison with Barton's suggestions}
In his 1982 paper~\cite{barton1982modelling}, Barton presented an attempt to extend the original Barton--Bandis model to \revised{cyclic} loading conditions.
Barton's attempt differs from the model extended here in the following four aspects.

\begin{itemize}\itemsep=0pt
  \item To incorporate position- and direction-dependence, Barton extended the $\jrc_{\rm{m}}$ evolution function (Table~\ref{tab:jrc-barton}) to the forward return, backward advance, and backward return stages. 
  Instead, our model has introduced the state-dependent variables ($\alpha$, $\Lambda_\mathrm{f}$, $\Lambda_\mathrm{b}$) for this purpose, keeping the same form of $\jrc_{\rm{m}}$ function originally proposed for the forward advance stage.

  \item In Barton's attempt, the (constant) shear stress in a return stage is larger than the residual shear stress. Conversely, in our model, the shear stress in a return stage is lower than the residual shear stress, as $\jrc_{\rm{m}}$ in Eq.~\eqref{eq:smooth-jrc-cyclic} is negative in a return stage ($\alpha=-1$).

  \item Barton neither considered contractive volume change in a return stage nor imposed a constraint on the amount of remaining dilation at the mated position. However, as in Eq.~\eqref{eq:dilation-angle-cyclic}, the dilation in our model is contractive in a return stage and designed to make the remaining dilation zero at the mated position.

  \item The magnitudes of $(\jrc_{p})_{\tau}$ in a backward stage, the case of $\delta<0$ in Eq.~\eqref{eq:jrcp-strength}, is different.
  Barton suggested that $(\jrc_{p})_{\tau}=0.75\jrc_{p}$, whereas we have adopted $(\jrc_{p})_{\tau}=0.87\jrc_{p}$ from Asadollahi and Tonon~\cite{asadollahi2011degradation}. 
\end{itemize}

Comparing the two versions of extension---one following Barton's suggestion~\cite{barton1982modelling} and the other described in the present work---with several sets of experimental data in the literature, we have found that the model extended in this work consistently shows better agreement with the experimental data.
This paper has thus proposed a new version of extension to cyclic loading.

\section{Implicit algorithm}
\label{sec:integration}
In this section, we introduce an algorithm for an implicit update of the extended Barton--Bandis model. 
The goal of the algorithm is as follows: given values at $t_{n}$ (the current time instance) and the displacement increment between $t_n$ and $t_{n+1}$ (the next time instance) $\Delta{\tensor{u}}$, find the traction vector $\bm{t}$ and the consistent tangent operator $\mathbb{C}$ at $t_{n+1}$.
Hereafter, we shall denote quantities at $t_{n}$ with subscript $(\cdot)_{n}$, and write quantities at $t_{n+1}$ without any subscript for brevity.

Leveraging the incremental elasto-plastic formulation of the Barton--Bandis model, we develop a return mapping procedure as in Algorithm~\ref{algo:material-update}. 
Similar to return mapping algorithms for other inelasticity models (\eg~\cite{borja2013plasticity,white2014anisotropic}), the algorithm uses a predictor--corrector approach that proceeds as follows:
\begin{enumerate}\itemsep=0pt
  \item Calculate a trial state assuming that the displacement increment is fully elastic.
  \item If the trial state is elastic ($F<0$), it is accepted as the final state.
  \item Otherwise, the trial state is corrected such that the final state satisfies constraints imposed by inelasticity.
\end{enumerate}
As is well known, this kind of return mapping algorithm has two main advantages over explicit algorithms: (i) it is unconditionally stable, and (ii) it guarantees satisfaction of the strength criterion ($F\leq 0$), without any overestimation of the strength.
In the following, we elaborate several points that are specific in the proposed return mapping algorithm for the Barton--Bandis model.
\begin{algorithm}[h!]
  \setstretch{1.25}
  \caption{Implicit update algorithm for the extended Barton--Bandis model.}
    \begin{algorithmic}[1]
      \Require The displacement increment $\Delta\tensor{u}$ at $t_{n+1}$.
      \State Calculate the total displacement, $\tensor{u} = \tensor{u}_{n} + \Delta \tensor{u}$, and decompose it into {$u_{\cn}$} and {$\delta$} as Eq.~\eqref{eq:u-decompose-direction}.
      \State Calculate the trial elastic displacement, $\tensor{u}^{\el,\trial} = \tensor{u}^{\el}_{n} + \Delta \tensor{u}$, and decompose it into {$u^{\el,\trial}_{\cn}$} and {$\delta^{\el,\trial}$} as Eq.~\eqref{eq:u-decompose-direction}.
      \State Calculate the trial normal stress, $\stress^{\trial}_{\cn}$, as Eq.~\eqref{eq:normal-stress} with ${u^{\el,\trial}_{\cn}}\rightarrow {u^{\el}_{\cn}}$.
      \State Calculate the trial shear stresses,  $\tau^{\trial}=\mu_{n} {\delta^{\el,\trial}}$. Here, $\mu_{n}$ is evaluated as in Eq.~\eqref{eq:shear-modulus} with $(\stress_{\cn})_{n}$. 
      \State Calculate the damage coefficient, $M$, as Eq.~\eqref{eq:damage-coeff} with $\stress^{\trial}_{\cn}\rightarrow \stress_{\cn}$. 
      \State Calculate the peak joint roughness coefficient for shear strength, $(\jrc_{{p}})_{\tau}$, as Eq.~\eqref{eq:jrcp-strength}.
      \State Calculate parameter $i_\tau$ as Eq.~\eqref{eq:parameter-i-cyclic}, with $\stress^{\trial}_{\cn}\rightarrow \stress_{\cn}$. 
      \State Calculate the mobilized joint roughness coefficient $\jrc_\mathrm{m}$ as Eq.~\eqref{eq:smooth-jrc-cyclic}. If ${\delta} \geq 0$, $(\Lambda_{\mathrm{f}})_{n} \rightarrow \Lambda$; otherwise, $(\Lambda_{\mathrm{b}})_{n} \rightarrow \Lambda$. 
      \State Calculate the friction angle $\phi$ as Eq.~\eqref{eq:friction-angle}, using $\jrc_\mathrm{m}$ computed above.
      \State Evaluate the yield function $F (\stress^{\trial}_{\cn}, \, \tau^\mathrm{tr}, \, \phi)$.
      \If {$F < 0$}
        \State Elastic joint displacement. 
        \State Update $\tensor{t} = \stress^{\trial}_{\cn} \tensor{n} + \tau^{\trial} \tensor{m}$, $\Lambda_\text{f} = (\Lambda_{\text{f}})_{n}$, and $\Lambda_\text{b} = (\Lambda_{\text{b}})_{n}$. 
        \State Calculate the elastic tangent operator $\mathbb{C}^{\el}$ as Eq.~\eqref{eq:elastic-tangent}, and let $\mathbb{C} = \mathbb{C}^{\el}$. 
      \Else 
        \State Inelastic joint displacement. 
        \State Update $\Lambda_\mathrm{f}$ and $\Lambda_\mathrm{b}$, as Eqs.~\eqref{eq:lambda-forward-update} and~\eqref{eq:lambda-backward-update}.
        \State Update $\jrc_\mathrm{m}$ as Eq.~\eqref{eq:smooth-jrc-cyclic}. If ${\delta} \geq 0$, $\Lambda_{\mathrm{f}} \rightarrow \Lambda$; otherwise, $\Lambda_{\mathrm{b}} \rightarrow \Lambda$. 
        \State Calculate the dilation angle $\psi$ as Eq.~\eqref{eq:dilation-angle-discrete}, with $\alpha=({\Delta \delta}/|{\Delta \delta}|)({\delta}/|{\delta}|)$.
        \State Solve for $\tensor{u}^{\el}$ and $\lambda$ using Newton's method, as Eqs.~\eqref{eq:return-mapping-unknown}--\eqref{eq:jacobian}.
        \State Calculate $\stress_{\cn}$ and $\tau$ using $\tensor{u}^{\el}$ computed above, and update $\tensor{t} = \stress_{\cn} \tensor{n} + \tau \tensor{m}$. 
        \State Calculate the elasto-plastic tangent operator $\mathbb{C}^\text{ep}$ as Eq.~\eqref{eq:elasto-plastic-tangent}, and let $\mathbb{C} = \mathbb{C}^\text{ep}$.
      \EndIf
      \Ensure $\tensor{t}$ and $\mathbb{C}$ at $t_{n+1}$.
    \end{algorithmic}
\label{algo:material-update}
\end{algorithm}

\paragraph{Type of shearing stage} 
To utilize the proposed model for a general cyclic loading problem, one must identify the type of the current shearing stage as Eq.~\eqref{eq:loading-stage}.
Note that the total slip displacement ${\delta}$ is used for this purpose.
In a discrete setting, $\alpha$ in Eq.~\eqref{eq:direction-alpha} can be calculated as ${(\Delta \delta/|\Delta \delta|)(\delta/|\delta|)}$, where ${\Delta \delta := \delta - \delta_n}$.

\paragraph{Stress-dependent shear stiffness} 
One challenge in implicit integration of the proposed elasto-plastic formulation is that the elastic shear stiffness $\mu$ is a function of the normal stress, see Eq.~\eqref{eq:shear-modulus}.
While the stress dependence of the shear stiffness can be incorporated in the algorithm, it gives rise to non-symmetry in the tangent operator, which may be undesirable for numerical performance.
Therefore, similar to how some return mapping algorithms have handled pressure-dependent elastic moduli in constitutive models for soils (\eg~\cite{borja1990cam}), here we evaluate the shear modulus using the normal stress at the current time instance, $t_n$. 
In this way, the elastic tangent operator is calculated as
\begin{align}
  \mathbb{C}^{\el} = K \tensor{n} \dyadic \tensor{n} + \mu_{n} \tensor{m} \dyadic \tensor{m} .
  \label{eq:elastic-tangent} 
\end{align}
where
\begin{align}
  K = \dfrac{\kappa (u^{\el}_{\max})^2}{(u^{\el}_{\max} - {u^{\el}_{\cn}})^2}, \quad
  \mu_{n} = \dfrac{(\stress_{\cn})_{n}\tan \phi_r}{0.3 {\delta_{p}}} . 
\end{align}

\paragraph{Accummulated slip displacements} 
The proposed model uses two variables for accumulated slip displacements, $\Lambda_{\mathrm{f}}$ in Eq.~\eqref{eq:lambda-forward} and $\Lambda_{\mathrm{b}}$ in~\eqref{eq:lambda-backward}, to distinguish between asperity damages in forward and backward stages, respectively.
To produce the initial shear strengths, both variables are initialized to be $0.3 \delta_p$, namely, $\Lambda_{\mathrm{f},0} = 0.3 {\delta_p}$ and $\Lambda_{\mathrm{b},0} = 0.3 {\delta_p}$,
where subscript $(\cdot)_0$ denotes the initial value of a time-dependent variable.
When inelastic deformation occurs, the two variables are updated as
\begin{align}
  \Lambda_\mathrm{f} &= (\Lambda_{\mathrm{f}})_{n} + H({\delta}) H(\alpha) \lvert {\Delta \delta} \rvert , \label{eq:lambda-forward-update} \\
  \Lambda_\mathrm{b} &= (\Lambda_{\mathrm{b}})_{n} + H({-\delta}) H(\alpha) \lvert {\Delta \delta} \rvert . \label{eq:lambda-backward-update} 
\end{align}
Once $\Lambda_\mathrm{f}$ and $\Lambda_\mathrm{b}$ are updated, $\Lambda$ is determined according to the current joint position (see Eq.~\eqref{eq:lambda-unified}) and then used to evaluate the mobilized joint roughness coefficient, $\jrc_{\mathrm{m}}$. 

\paragraph{Dilation angle} 
As in Eq.~\eqref{eq:dilation-angle-cyclic}, the proposed model uses two expressions for the dilation angle to produce dilation in an advance stage (as in the original Barton--Bandis model) and contraction in an return stage.
The dilation angle in the return stage is calculated such that it gives zero volume change when the joint goes back to the initial position.
Therefore, the dilation angle at the next time instance should be calculated with the remaining amount of dilation and slip displacement at the current time instance. \revised{Otherwise, the model may show over-contraction at the initial position.}
So Eq.~\eqref{eq:dilation-angle-cyclic} is evaluated as follows:
\begin{align}
  \psi = 
  \begin{cases}
    \dfrac{1}{M} \jrc_\mathrm{m} \log\left(\dfrac{\jcs}{\stress_{\cn}} \right) & \text{if}\; \alpha = +1 , \vspace{2em} \\
    - \arctan \left(\dfrac{(u^{\pl}_{\cn})_{n}}{ \lvert {\delta_{n}} \rvert} \right) \dfrac{180}{\pi} & \text{if}\; \alpha = -1,
  \end{cases}
  \quad \text{(in degrees).}
  \label{eq:dilation-angle-discrete}
\end{align}
Note that the first equation is calculated with quantities at $t_{n+1}$ \revised{for fully implicit integration in the loading phase, whereas the second one is evaluated with those at $t_{n}$ to ensure zero volume change when the joint is returned to the initial position}.

\paragraph{Newton's method for inelastic correction} 
In the corrector step, Newton's method is used to solve for the elastic displacement vector, $\tensor{u}^{\el}$, and the discrete plastic multiplier, $\Delta\lambda$. 
The unknown vector can be written as 
\begin{align}
  \tensor{x} = 
  \begin{bmatrix}
    (\tensor{u}^{\el})_{3 \times 1} \\ 
    \Delta \lambda
  \end{bmatrix}_{4 \times 1} . 
  \label{eq:return-mapping-unknown}
\end{align}
The residual vector is composed of four equations that need to be satisfied, \ie 
\begin{align}
  \tensor{r}(\bm{x}) = 
  \begin{bmatrix}
    \left(\tensor{u}^{\el} - \tensor{u}^{\el,\trial} + \Delta \lambda \dfrac{\pd G}{\pd \tensor{t}} \right)_{3 \times 1} \\ 
    F
  \end{bmatrix}_{4 \times 1} \rightarrow \tensor{0} .
  \label{eq:return-mapping-residual}
\end{align}
At each Newton iteration, the unknown vector is updated by solving
\begin{align}
  \tensor{J} \cdot \Delta \tensor{x} = - \tensor{r}(\bm{x}),  
  \label{eq:return-mapping-newton}
\end{align}
where the Jacobian matrix $\tensor{J}$, given by 
\begin{align}
  \tensor{J} = 
  \begin{bmatrix}
    \left(\tensor{1} + \Delta \lambda \dfrac{\pd^{2} G}{\pd \tensor{t} \dyadic \pd \tensor{t}} \cdot \mathbb{C}^{\el} \right)_{3 \times 3} & \left(\dfrac{\pd G}{\pd \tensor{t}}\right)_{3 \times 1} \vspace{1em} \\
    \left( \dfrac{\pd F}{\pd \tensor{t}} \cdot \mathbb{C}^{\el} \right)^{\intercal}_{1 \times 3} & 0
  \end{bmatrix}_{4 \times 4} , 
  \label{eq:jacobian}
\end{align}
with $\tensor{1}$ denoting the second-order identity tensor. 
Specific expressions for the derivatives of the yield function $F$ and the potential function $G$ are provided in~\ref{sec:appendix}. 
It is noted that we have used the fact that the friction and dilation angles of the Barton--Bandis model are independent of the plastic multiplier.
Note also that at the end of each iteration, all variables related to the elastic displacement---the traction vector and other stress-dependent variables---should be updated.

\paragraph{Consistent tangent operator} 
To use the model in a stress-controlled problem as well as in an implicit numerical method, the consistent tangent operator should be calculated. 
Depending on whether the deformation is fully elastic or not, the consistent tangent operator takes different forms as
\begin{equation}
  \mathbb{C} = \dfrac{\pd \tensor{t}}{\pd \tensor{u}^{\el, \trial}} =
  \begin{cases}
    \mathbb{C}^{\el} & \text{if}\; F < 0, \\
    \mathbb{C}^\text{ep} & \text{if}\; F = 0. \\
  \end{cases}
\end{equation}
The elastic tangent $\mathbb{C}^{\el}$ can be calculated as Eq.~\eqref{eq:elastic-tangent},
and the elasto-plastic tangent $\mathbb{C}^{\text{ep}}$ can be derived as follows.  
First, we linearize Eq.~\eqref{eq:return-mapping-residual} with respect to $\tensor{u}^{\el, \trial}$ and obtain 
\begin{align}
  \begin{bmatrix}
    \left( \revised{\mathcal{L} ( \tensor{u}^{\el})}  + \Delta \lambda \dfrac{\pd^2 G}{\pd \tensor{t} \dyadic \pd \tensor{t}} \cdot \mathbb{C}^{\el} \cdot \revised{\mathcal{L} ( \tensor{u}^{\el})} + \dfrac{\pd G}{\pd \tensor{t}} \revised{\mathcal{L} ( \Delta \lambda)}   \right)_{3 \times 1} \\
    \dfrac{\pd F}{\pd \tensor{t}} \cdot \mathbb{C}^{\el} \cdot \revised{\mathcal{L} ( \tensor{u}^{\el})} 
  \end{bmatrix}_{4 \times 1} = 
  \begin{bmatrix}
    \left(\revised{\mathcal{L} ( \tensor{u}^{\el, \trial})} \right)_{3 \times 1} \\ 
    0
  \end{bmatrix}_{4 \times 1} . 
  \label{eq:relinearize-residual}
\end{align}
where \revised{$\mathcal{L}(\cdot)$} denotes the linearization operator.
Inserting $\revised{\mathcal{L} ( \tensor{u}^{\el})} = (\mathbb{C}^{\el})^{-1} \cdot \revised{\mathcal{L} ( \tensor{t})} $ into the above equation, we obtain 
\begin{align}
  \left[(\mathbb{C}^{\el})^{-1} + \Delta \lambda \frac{\pd^2 G}{\pd \tensor{t} \dyadic \pd \tensor{t}}  \right] \cdot \revised{\mathcal{L} ( \tensor{t} )} + \dfrac{\pd G}{\pd \tensor{t}} \revised{\mathcal{L} ( \Delta \lambda )} &= \revised{\mathcal{L} ( \tensor{u}^{\el, \trial})} , \label{eq:relinearize-flow-rule-eq} \\ 
  \dfrac{\pd F}{\pd \tensor{t}} \cdot \revised{\mathcal{L} ( \tensor{t} )} &= 0 .  \label{eq:relinearize-consistency-eq}
\end{align}
Then we rearrange Eq.~\eqref{eq:relinearize-flow-rule-eq}, which gives 
\begin{align}
  \revised{\mathcal{L} ( \tensor{t})} = \mathbb{P} \cdot \left(\revised{\mathcal{L} ( \tensor{u}^{\el, \trial} )} -  \dfrac{\pd G}{\pd \tensor{t}}   \revised{\mathcal{L} ( \Delta \lambda )} \right), \label{eq:rearranged-flow-rule}
\end{align}
where $\mathbb{P}$ is defined as 
\begin{align}
  \mathbb{P} = \left(\tensor{1} +   \Delta \lambda \frac{\pd^2 G}{\pd \tensor{t} \dyadic \pd \tensor{t}} \cdot \mathbb{C}^{\el} \right)^{-1} \cdot \mathbb{C}^{\el}. 
\end{align}
By differentiating the two sides of Eq.~\eqref{eq:rearranged-flow-rule} with respect to $\tensor{u}^{\el, \trial}$, we get $\mathbb{C}^{\text{ep}}$ as
\begin{align}
  \mathbb{C}^{\text{ep}} = \dfrac{\pd \tensor{t}}{\pd \tensor{u}^{\el, \trial}} = \mathbb{P} \cdot \left(\tensor{1} -  \dfrac{\pd G}{\pd \tensor{t}}  \dfrac{\pd \Delta \lambda}{\pd \tensor{u}^{\el, \trial}} \right) . \label{eq:cto-ep}
\end{align}
The only unknown in Eq.~\eqref{eq:cto-ep} is $\pd \Delta \lambda / \pd \tensor{u}^{\el, \trial}$. 
To evaluate this, we insert Eq.~\eqref{eq:rearranged-flow-rule} into Eq.~\eqref{eq:relinearize-consistency-eq} and obtain
\begin{align}
  \dfrac{\pd F}{\pd \tensor{t}} \cdot \mathbb{P} \cdot \left(\revised{\mathcal{L} ( \tensor{u}^{\el, \trial})} -  \dfrac{\pd G}{\pd \tensor{t}} \revised{\mathcal{L} ( \Delta \lambda )} \right) = 0 . \label{eq:relinearize-consistency-new}
\end{align}
Rearranging the above equation and differentiating it with respect to $\tensor{u}^{\el,\trial}$ gives
\begin{align}
  \dfrac{\pd \Delta \lambda}{\pd \tensor{u}^{\el,\trial}} = \dfrac{\dfrac{\pd F}{\pd \tensor{t}} \cdot \mathbb{P}}{\dfrac{\pd F}{\pd \tensor{t}} \cdot \mathbb{P}  \cdot \dfrac{\pd G}{\pd \tensor{t}} } . 
\end{align}
Finally, inserting the above equation into Eq.~\eqref{eq:cto-ep}, we get
\begin{align}
  \mathbb{C}^\text{ep} = \mathbb{P} - \dfrac{\left(\mathbb{P} \cdot \dfrac{\pd G}{\pd \tensor{t}}  \right) \otimes \left( \dfrac{\pd F}{\pd \tensor{t}} \cdot \mathbb{P}\right)}{\dfrac{\pd F}{\pd \tensor{t}} \cdot \mathbb{P} \cdot \dfrac{\pd G}{\pd \tensor{t}} } . 
  \label{eq:elasto-plastic-tangent}
\end{align}
\smallskip

\begin{remark}
  The implicit algorithm developed in this work (Algorithm~\ref{algo:material-update}) can also be used for the original Barton--Bandis model, by specializing it to a forward advance stage ($\delta\geq 0$ and $\alpha=1$).
\end{remark}

\section{Verification and validation}
\label{sec:numerical}

This section has two objectives: (i) to verify the elasto-plastic formulation of the Barton--Bandis model and the implicit integration algorithm, and (ii) to validate the extension of the formulation to cyclic loading.
For the first purpose, we simulate the monotonic shear tests modeled by Barton \etal~\cite{barton1985strength} and compare the results of the elasto-plastic formulation and the original one.
For the second purpose, we apply the extended Barton--Bandis model to simulate the cyclic shear tests of Lee~\etal~\cite{lee2001influence} conducted on four rock joint samples.

\subsection{Verification against the original Barton--Bandis model under monotonic loading}
For verification, we use the proposed elasto-plastic formulation to simulate two sets of monotonic shear box tests modeled by Barton~\etal~\cite{barton1985strength}. 
The first set demonstrates how the model captures the stress dependence of rock joint behavior, by shearing a 0.3-m long rock joint under three different normal stresses, namely $3$ MPa, $10$ MPa and $30$ MPa. 
The second set is concerned with size (scale) effects on rock joint behavior, simulating three rock joints of different sizes, namely 0.1 m, 1 m, and 2 m, sheared under the same normal stress of 2 MPa.
Table~\ref{tab:verification-parameters} presents the material parameters used in the two sets, which are adopted from Barton~\etal~\cite{barton1985strength}.
To keep the normal stress constant during shearing, we make use of a global Newton iteration similar to how to solve a mixed boundary-value problem. 
Being consistent with Barton~\etal~\cite{barton1985strength}, the damage coefficient is set to be $M = 2$ for all the cases.
\begin{table}[h!]
    \centering
    \begin{tabular}{l|l|l|l}
    \toprule
    \multirow{2}{*}{Parameter} & \multirow{2}{*}{Unit}  & \multicolumn{2}{c}{Value}  \\
    \cline{3-4}
    & & Stress effects (Fig.~\ref{fig:monotonic-barton-stress}) & Size effects (Fig.~\ref{fig:monotonic-barton-scale}) \\  
    \midrule
    $\phi_{r}$ & degrees & 30.0 & 30.0\\
    $\jrc_{p0}$ & - & 10.0 & 15.0 \\
    $\jcs_{0}$ & MPa & 100.0 & 150.0 \\
    $l_0$ & m & 0.1  & 0.1 \\ 
    \bottomrule
    \end{tabular}
    \caption{Material parameters for the two sets of verification examples, adopted from Barton \etal~\cite{barton1985strength}.}
    \label{tab:verification-parameters}
\end{table}

\begin{figure}[h!]
    \centering
    \subfloat[Shear stress]{\includegraphics[width=0.48\textwidth]{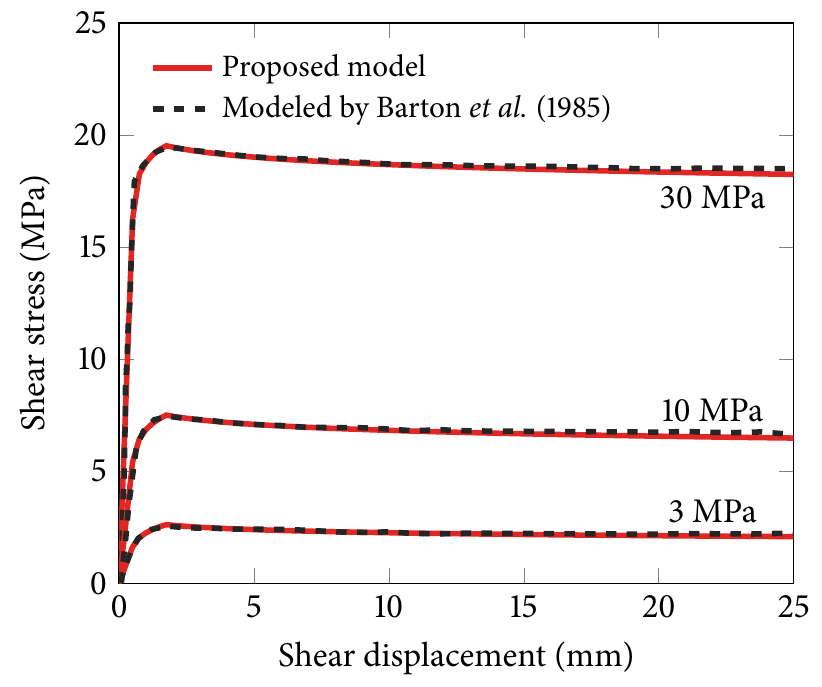}} \hspace{0.5em}
    \subfloat[Dilation]{\includegraphics[width=0.48\textwidth]{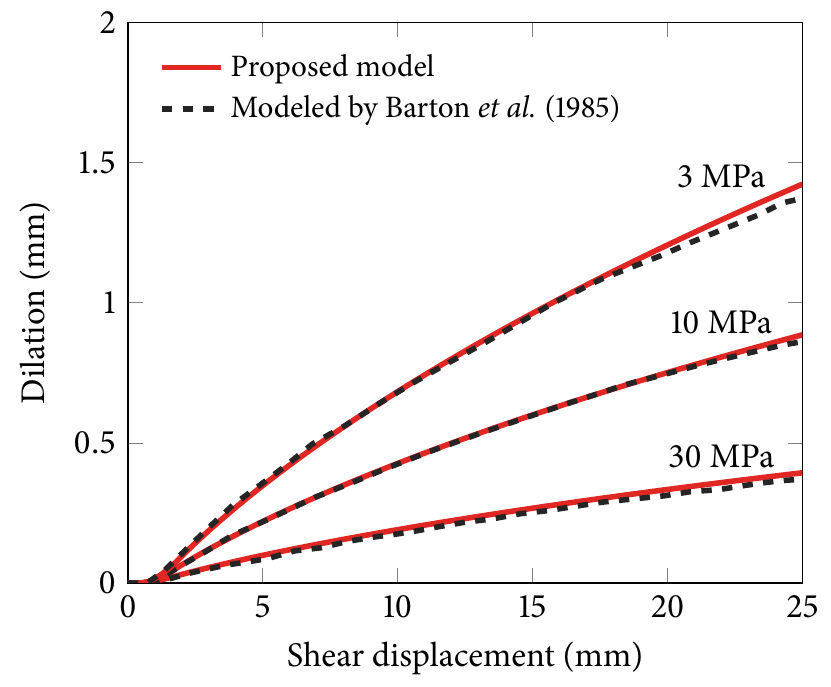}}
    \caption{Verification of the proposed model under different normal stresses: (a) shear stress and (b) dilation compared with the results of the original Barton--Bandis model in Barton \etal~\cite{barton1985strength}.}
    \label{fig:monotonic-barton-stress}
\end{figure}

Figure~\ref{fig:monotonic-barton-stress} compares the simulation results for the first set ($\sigma_{\cn}$ = 3 MPa, 10 MPa, and 30 MPa) obtained by the elasto-plastic formulation and the original version~\cite{barton1985strength}.
Under all the three normal stresses, the shear stress and dilation responses simulated by the proposed model are virtually the same as those by the original model.  
The results also demonstrate that the Barton--Bandis model can capture the effects of normal stress on the shear strength, stiffness, and the amount of dilation.
Remarkably, all the initial slopes of the shear stress--displacement curves agree well with each other. This agreement indicates that Eq.~\eqref{eq:shear-modulus} incorporates the stress dependence of elastic shear stiffness as implied in the Barton--Bandis model. 

\begin{figure}[h!]
    \centering
    \subfloat[Shear stress]{\includegraphics[width=0.48\textwidth]{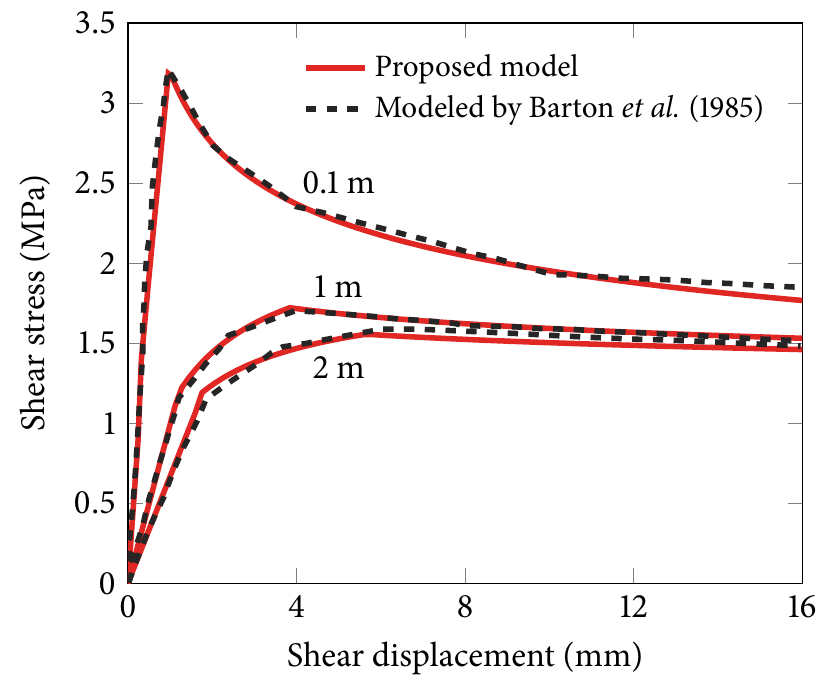}} \hspace{0.5em}
    \subfloat[Dilation]{\includegraphics[width=0.48\textwidth]{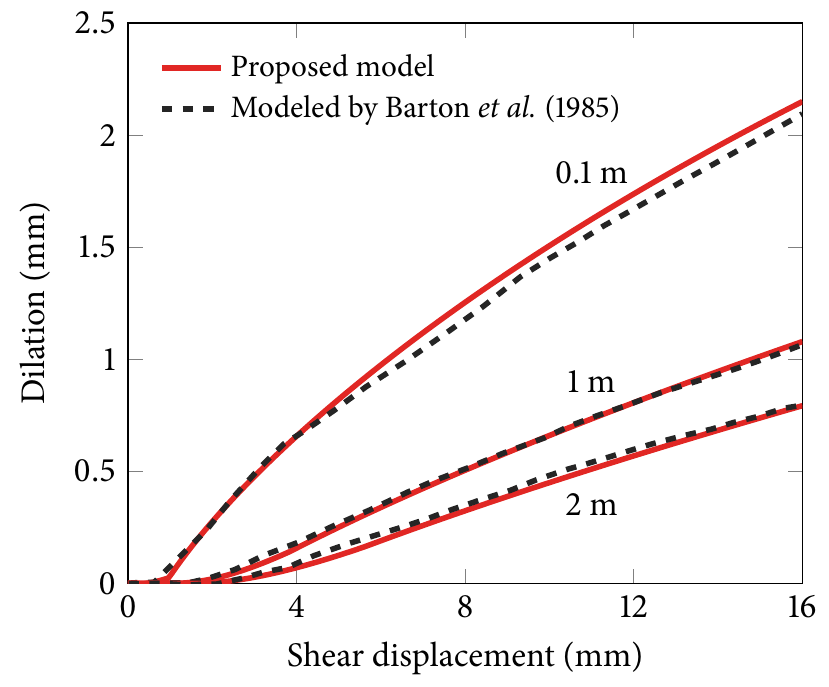}}
    \caption{Verification of the proposed model for joints with different joint sizes: (a) shear stress and (b) dilation compared with the results of the original Barton--Bandis model in Barton \etal~\cite{barton1985strength}.}
    \label{fig:monotonic-barton-scale}
\end{figure}

Next, in Fig.~\ref{fig:monotonic-barton-scale} we compare the two models' simulation results for the second set ($l_j$ = 0.1 m, 1 m, and 2 m).
It can be seen that the proposed model can also capture size (scale) effects in the same way as in the original Barton--Bandis model.
The initial slopes of the two models again show excellent agreement, indicating that Eq.~\eqref{eq:shear-modulus} correctly incorporates size effects on the elastic shear stiffness (through the size dependence of $\delta_p$).
It further suggests that the slight difference between the two results at large shear displacements are due to the algorithm, rather than the proposed formulation for the elastic shear stiffness.
\revised{It is noted that the implicit integration algorithm enforces the strength criterion to be satisfied in each step, whereas an explicit integration algorithm does not. As such, the result obtained by the implicit algorithm can be considered more consistent with the strength criterion.}
Taken together, it has been verified that the proposed elasto-plastic formulation inherits the capabilities of the original Barton--Bandis model.

\begin{figure}[h!]
    \centering
    \subfloat[Local convergence]{\includegraphics[width=0.48\textwidth]{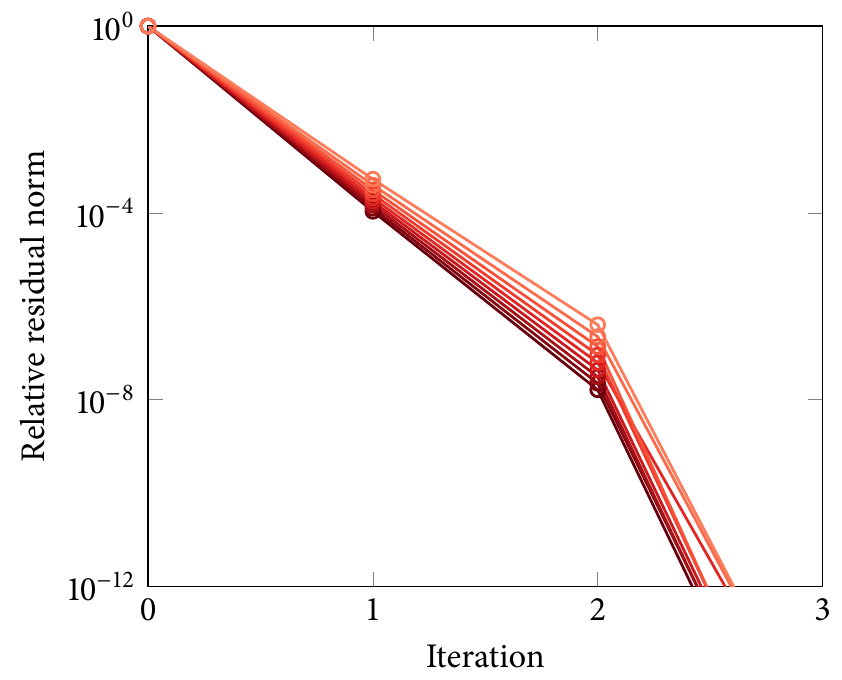} \label{fig:local-convergence}}   \hspace{0.5em}
    \subfloat[Global convergence]{\includegraphics[width=0.48\textwidth]{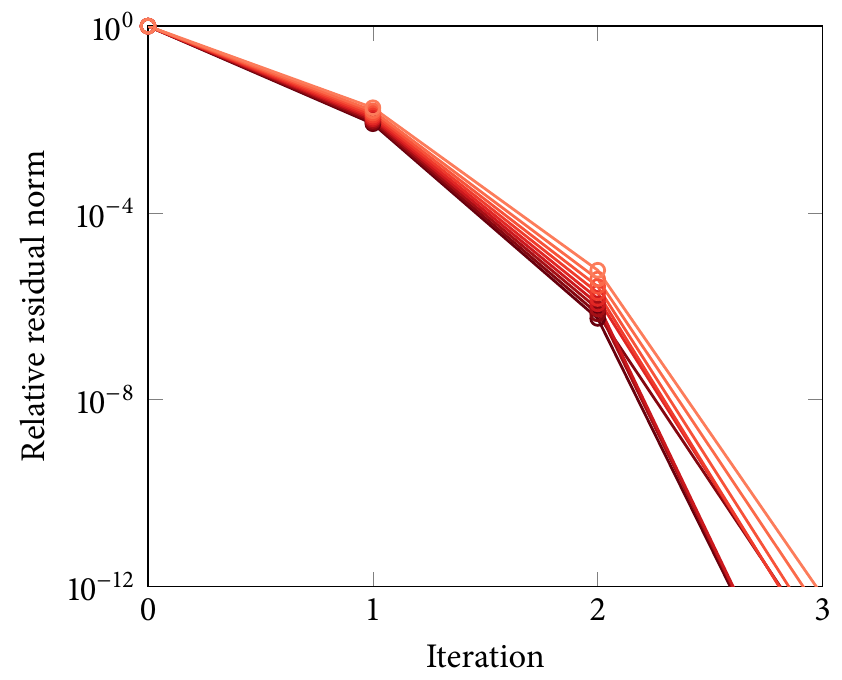} \label{fig:global-convergence}} 
    \caption{Verification of the proposed model: Newton convergence profiles during (a) local return mapping and (b) global stress control.}
    \label{fig:newton-convergence}
\end{figure}

Lastly, in Fig.~\ref{fig:newton-convergence} we present typical Newton convergence profiles in local return mapping and global stress control for the foregoing simulations.
As shown in Fig.~\ref{fig:local-convergence}, the Newton iterations during local return mapping displays asymptotically quadratic convergence, which affirms the correctness of the Jacobian matrix~\eqref{eq:jacobian}.
The global convergence behavior shown in Fig.~\ref{fig:global-convergence} also exhibits nearly quadratic rates, which verifies the elasto-plastic tangent operator~\eqref{eq:elasto-plastic-tangent}. 
These results indicate that the proposed algorithm allows one to use the Barton--Bandis model with high robustness and efficiency.

\subsection{Validation against experimental data on rock joints under cyclic loading}
Having verified our formulation and algorithm, we validate the extended Barton--Bandis model against the responses of real rock joints measured in cyclic shear box tests. 
Particularly, we use the experimental data on rock joint samples from Hwangdeung granite and Yeosan marble in Lee~\etal~\cite{lee2001influence}.
We choose four joint samples named GH18, GH27, GH45, and MH34 therein. 
The first three joints are from Hwangdeung granite joints and the last one from Yeosan marble.
The normal stress was 1 MPa for the GH18 and GH27 samples and 3 MPa for the GH45 and MH34 samples.  
Table~\ref{tab:validation-parameters} presents the material parameters of the four joint samples. 
All the values are adopted from Lee~\etal~\cite{lee2001influence}, except for $\jrc_{p0}$ whose values are calibrated to match the experimental data. 
\begin{table}[h!]
    \centering
    \begin{tabular}{l|l|m{14mm}|m{14mm}|m{14mm}|m{14mm}}
    \toprule
    \multirow{2}{*}{Parameter} & \multirow{2}{*}{Unit} & \multicolumn{4}{c}{Value} \\
    \cline{3-6} 
    & &  Granite GH18 & Granite GH27 & Granite GH45 & Marble MH34 \\ 
    \midrule
    $\phi_{r}$ & degrees  & 34.6 & 34.6 & 34.6 & 38.3 \\
    $\jrc_{p0}$ & -  & 9.0 & 7.8 & 9.0 & 13.0 \\
    $\jcs_{0}$ & MPa & 151.0 & 151.0 & 151.0 & 72.0 \\
    $l_{0}$ & m  & 0.12  & 0.12 & 0.12 & 0.12 \\ 
    $l_{j}$ & m  & 0.12 & 0.12 & 0.12 & 0.12 \\
    \bottomrule
    \end{tabular}
    \caption{Material parameters for the validation examples. All the values except $\jrc_{p0}$ are adopted from Lee \etal~\cite{lee2001influence}.}
    \label{tab:validation-parameters}
\end{table}

\begin{figure}[h!]
  \centering
  \subfloat[Shear stress]{\includegraphics[width=0.48\textwidth]{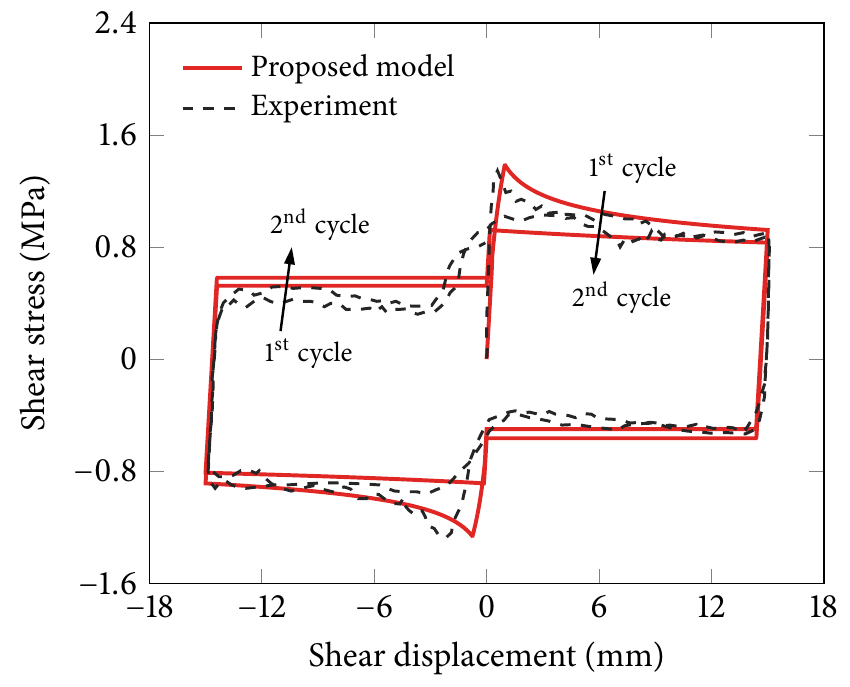}} \hspace{0.5em}
  \subfloat[Dilation]{\includegraphics[width=0.48\textwidth]{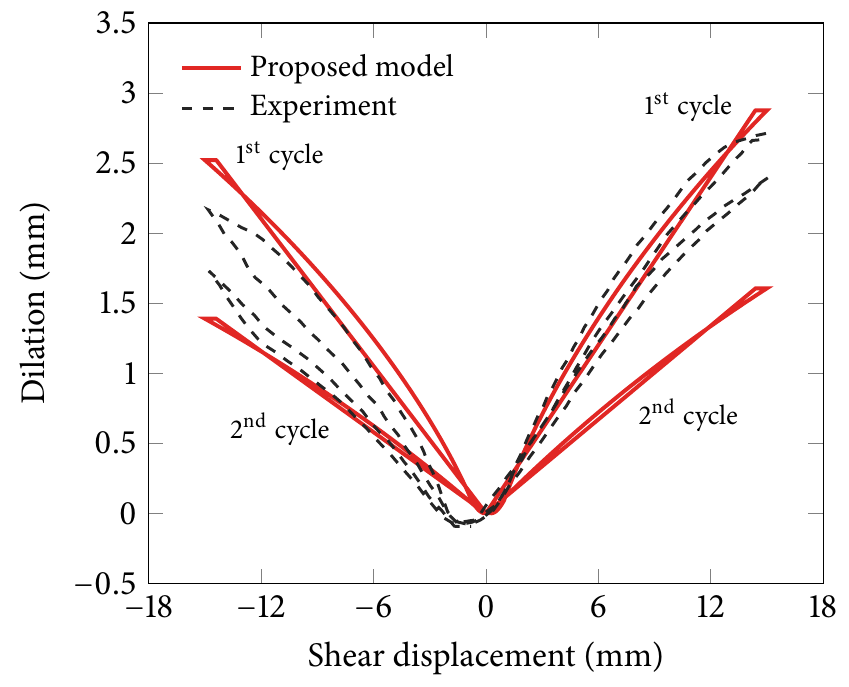}}
  \caption{Validation of the proposed model: (a) shear stress and (b) dilation compared with the experimental results on the GH18 joint in Lee \etal~\cite{lee2001influence}.}
  \label{fig:cyclic-lee-granite-1mpa}
\end{figure}

\begin{figure}[h!]
    \centering
    \subfloat[Shear stress]{\includegraphics[width=0.48\textwidth]{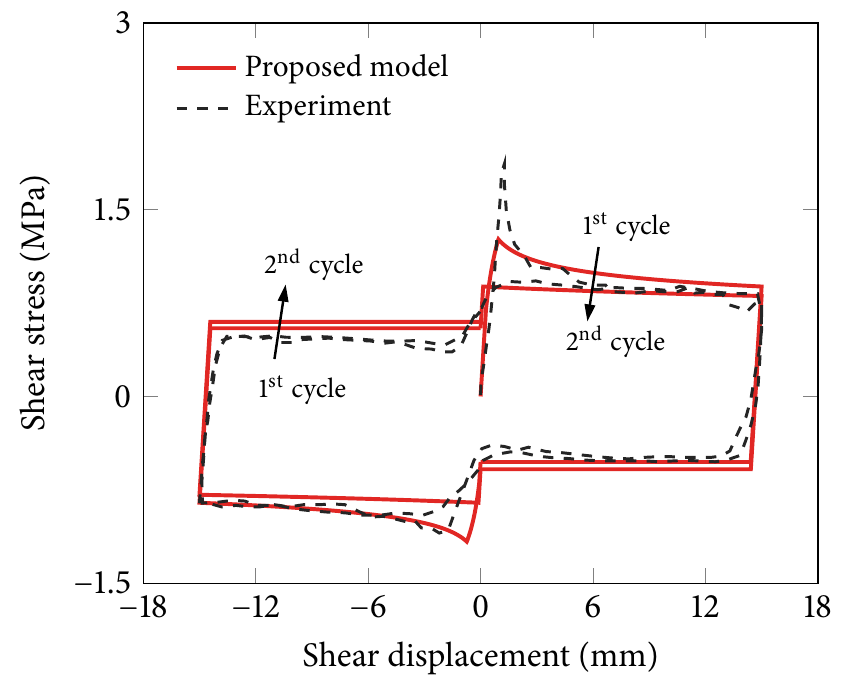}} \hspace{0.5em}
    \subfloat[Dilation]{\includegraphics[width=0.48\textwidth]{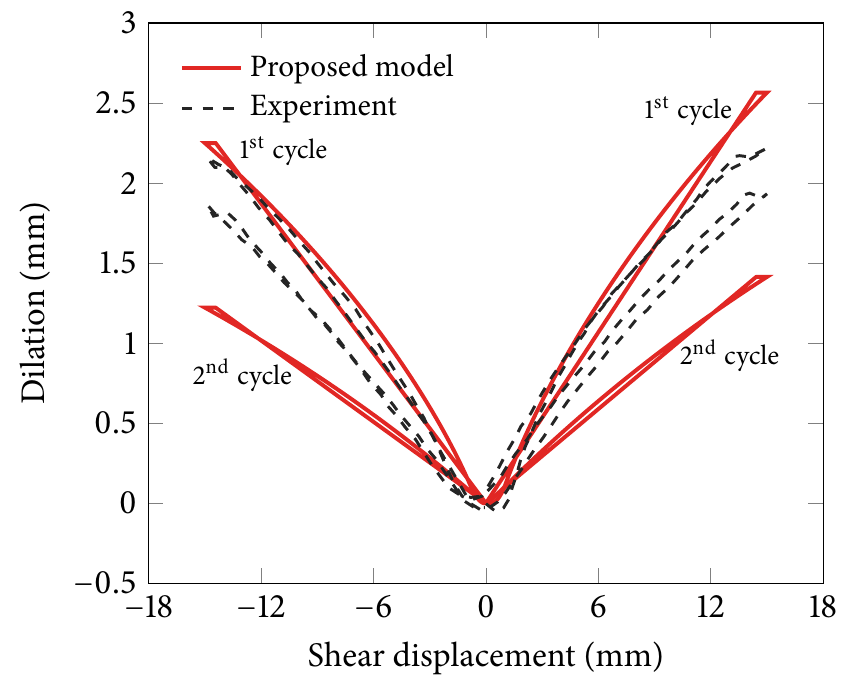}}
    \caption{Validation of the proposed model: (a) shear stress and (b) dilation compared with the experimental results on the GH27 joint in Lee \etal~\cite{lee2001influence}.}
    \label{fig:cyclic-lee-granite-1mpa-2}
\end{figure}

\begin{figure}[h!]
  \centering
  \subfloat[Shear stress]{\includegraphics[width=0.48\textwidth]{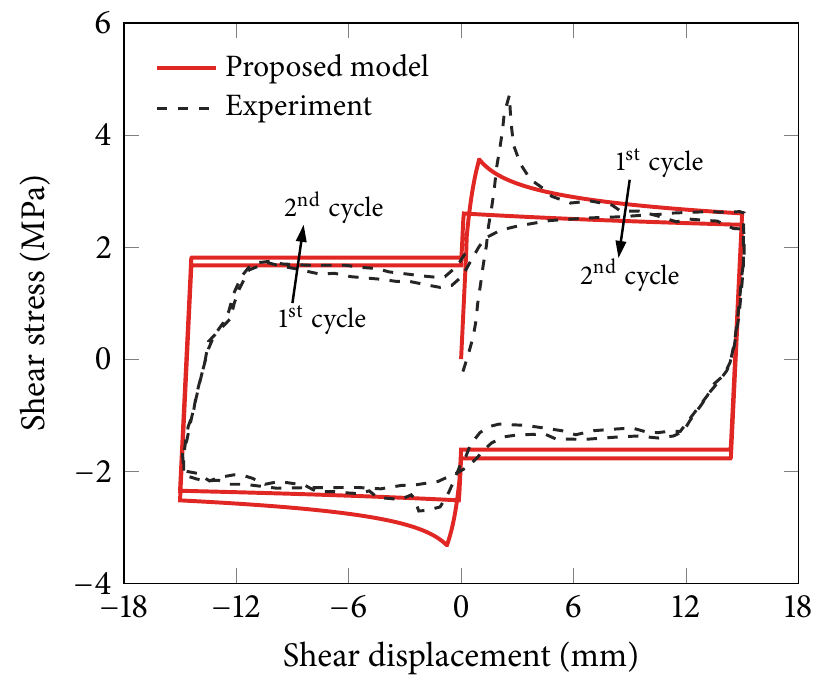}} \hspace{0.5em}
  \subfloat[Dilation]{\includegraphics[width=0.48\textwidth]{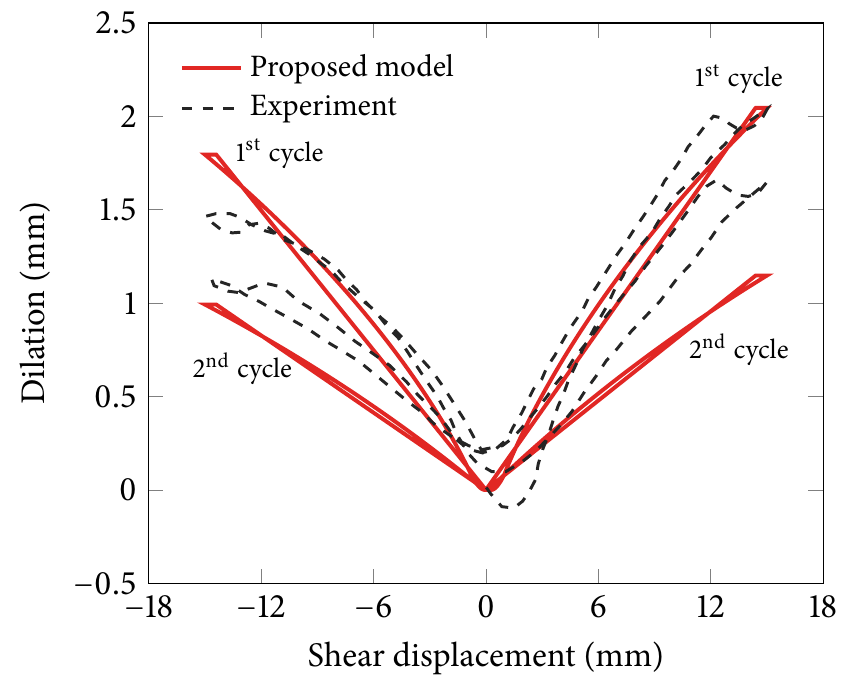}}
  \caption{Validation of the proposed model: (a) shear stress and (b) dilation compared with the experimental results on the GH45 joint in Lee \etal~\cite{lee2001influence}.}
  \label{fig:cyclic-lee-granite-3mpa}
\end{figure}

\begin{figure}[h!]
  \centering
  \subfloat[Shear stress]{\includegraphics[width=0.48\textwidth]{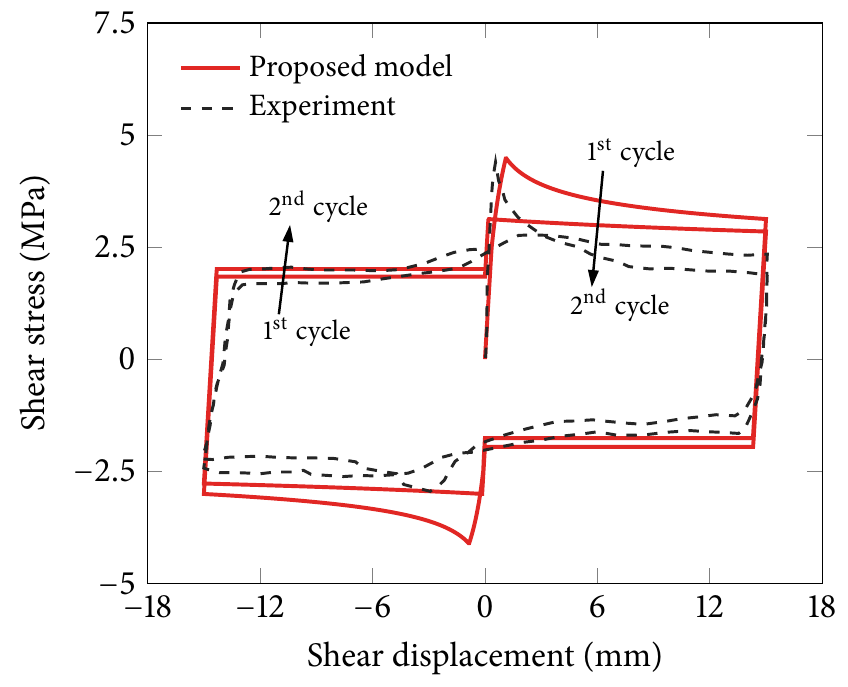}} \hspace{0.5em}
  \subfloat[Dilation]{\includegraphics[width=0.48\textwidth]{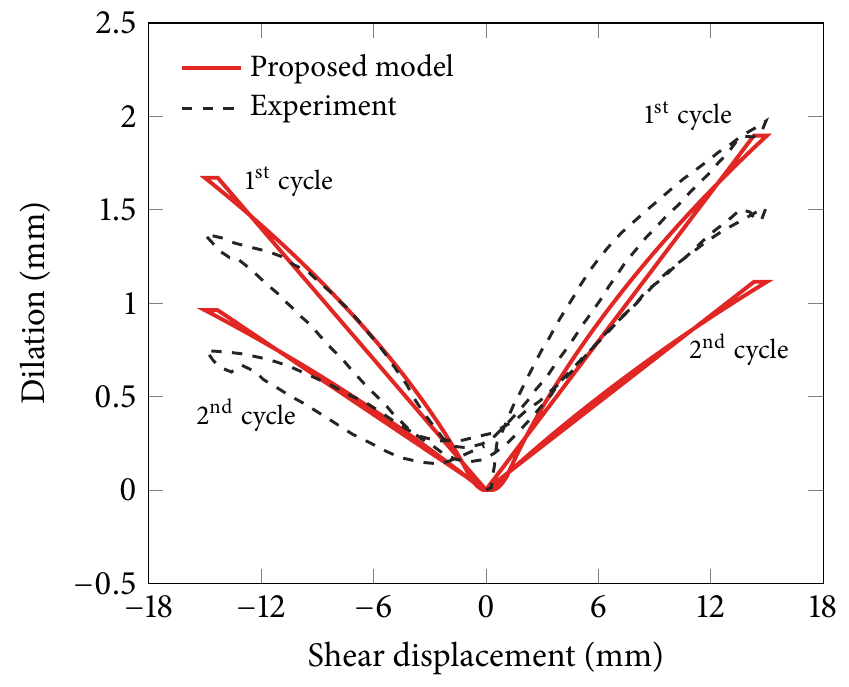}}
  \caption{Validation of the proposed model: (a) shear stress and (b) dilation compared with the experimental results on the MH34 joint in Lee \etal~\cite{lee2001influence}.}
  \label{fig:cyclic-lee-marble-3mpa}
\end{figure}

Figures~\ref{fig:cyclic-lee-granite-1mpa}--\ref{fig:cyclic-lee-marble-3mpa} compare the simulation results and experimental data of the four joint samples.
It can be seen that the simulation results show excellent qualitative agreement with the experimental data.
The quantitative agreement is also satisfactory, considering that no additional parameter is introduced for cyclic loading.
It is noted that, while not presented for brevity, we have confirmed that the extension proposed in this work match the experimental data better than the attempt made in Barton~\cite{barton1982modelling}.
Detailed discussions on the model behavior in each shearing stage are provided below.
\begin{itemize}\itemsep=0pt
  \item  \textbf{Forward advance.} 
  The model behavior in the first load cycle is the same as the original Barton--Bandis model, as demonstrated earlier in Figs.~\ref{fig:monotonic-barton-stress} and~\ref{fig:monotonic-barton-scale}.
  However, after asperities have been damaged in the first cycle, the model shows less dilation and thus a lower peak stress. 
  This hysteresis is consistent with the experimental observations.
  \revised{Yet the experimental and simulations results have a couple of notable differences.
  Compared with the experimental results, the simulation results have a smaller stress drop in the post-peak stage and less dilation in the second cycle.
  These differences, which are associated with the original Barton--Bandis model, may be attributed to the fact that the model only considers a single order of asperities characterized by $\jrc$.
  In real rock joints, asperities with different scales and degradation rates may coexist on a single joint surface~\cite{lee2001influence,li2019predicting}, and they may result in a larger post-peak stress drop and a higher residual dilation angle. 
  }

  \item \textbf{Forward return.}
  During the elastic unloading phase, the slope of the simulated shear stress curve matches well with the experimental data. This indicates that the elastic shear modulus derived as Eq.~\eqref{eq:shear-modulus} is not only consistent with the original Barton--Bandis model but also physically realistic.
  In the following phase of reversed shearing, the model produces a constant shear stress and a contractive normal displacement such that no volume change remains when the joint returns to the initial mated position.
  The shear stress in the second load cycle is slightly lower than that in the first cycle. 
  All these model behaviors are consistent with the experimental observations.

  \item \textbf{Backward advance.} 
  When the joint position is backward, the model gives a lower peak strength and less dilation than those in the forward position.
  The peak strength and dilation become further reduced in the second load cycle. 
  So it can be seen that the model captures not only asperity damage during cyclic loading but also its dependence on the direction of shear displacement.

  \item \textbf{Backward return.} 
  Similar to the forward return stage, the model produces a constant shear stress following a short period of elastic unloading. 
  The shear strength in the backward position is lower than that in the forward position. 
  Also, the amount of contraction is lower in the second cycle, because less dilation has taken place in the backward advance stage of the second cycle.
  As a result of this less contraction, the shear stress slightly increases in the second cycle. 
  Note that the same difference also exists in the experimental data.
\end{itemize}


To conclude, the extended Barton--Bandis model can reproduce all the salient features of rock joint behavior under cyclic loading, without any free parameter introduced to the original model.
Therefore, the proposed model is believed to be one of the most capable and practical means to simulate and predict the behavior of rock joints under cyclic loading.

\section{Closure}
\label{sec:closure}

In this paper, we have extended the Barton--Bandis model to rock joints under cyclic loading conditions, developing an algorithm for the robust and accurate use of the model in numerical simulation.
The extended model is free of any new material parameter and equivalent to the original model under monotonic shearing conditions.
As such, the main features and practical merits of the Barton--Bandis model have been carried over to cyclic loading conditions.
Also, the implicit algorithm developed herein enables one to use the Barton--Bandis model, in both original and extended forms, to be compatible with state-of-the-art numerical methods for fracture propagation and/or coupled multiphysical problems (\eg~\cite{settgast2017fully,choo2018cracking,choo2019stabilized,fei2020phaseb,lepillier2020variational,fei2021double,kadeethum2021locally}).
The contributions of this work will thus help address a large number of rock joint problems in research and practice. 
\revised{Examples range from the stability of jointed rock masses under periodic drilling and blasting in infrastructure engineering and mining operations, to cyclic stimulation of deep subsurface systems in modern energy technologies.}

\section*{Acknowledgments}
The authors wish to express their deep gratitude to Dr. Nick Barton for his careful review of the manuscript and constructive suggestions.
This work was supported by the National Research Foundation of Korea (NRF) grant funded by the Korean government (MSIT) (No. 2022R1F1A1065418).
Portions of this work were performed under the auspices of the U.S. Department of Energy by Lawrence Livermore National Laboratory under Contract DE-AC52-07NA27344.

\appendix

\section{Derivatives in the implicit algorithm}
\label{sec:appendix}
This appendix provides specific expressions for the derivatives in the implicit solution algorithm.
First, the derivatives of the yield function~\eqref{eq:yield-function} and the potential function~\eqref{eq:potential-function} are given by
\begin{align}
  \dfrac{\pd F}{\pd \tensor{t}} &= \tensor{m} + \tensor{n} \tan \phi + \stress_{\cn} \dfrac{\pd \tan \phi}{\pd \stress_{\cn}} \tensor{n},  \label{eq:grad-F}  \\  
  \dfrac{\pd G}{\pd \tensor{t}} &= \tensor{m} + \tensor{n}\tan \psi\\
  \dfrac{\pd^{2} G}{\pd \tensor{t} \dyadic \pd \tensor{t}} &= \dfrac{\pd \tan \psi}{\pd \stress_{\cn}} \tensor{n} \dyadic \tensor{n} ,
\end{align}
The derivatives of $\phi$ and $\psi$ with respect to $\sigma_{\cn}$, respectively, are given by
\begin{align}
  \dfrac{\pd \tan \phi}{\pd \stress_{\cn}} = (1 + \tan^2 \phi) \left[ -\dfrac{\jrc_\mathrm{m}}{ \stress_{\cn} \ln 10 } + \log \left(\dfrac{\jcs}{\stress_{\cn}} \right)\dfrac{\pd \jrc_\mathrm{m}}{\pd \stress_{\cn}} \right] ,  
  \label{eq:friction-deriv}
\end{align}
and
\begin{align}
  \dfrac{\pd \tan \psi}{\pd \stress_{\cn}} = 
  \begin{cases}
  \dfrac{1 + \tan^2 \psi}{M^2} \left[ - \dfrac{\jrc_\mathrm{m}M}{ \stress_{\cn} \ln 10 } + M \log \left(\dfrac{\jcs}{\stress_{\cn}} \right)\dfrac{\pd \jrc_\mathrm{m}}{\pd \stress_{\cn}} - \jrc_\mathrm{m} \log \left(\dfrac{\jcs}{\stress_{\cn}} \right) \dfrac{\pd M}{\pd \stress_{\cn}} \right] & \text{if}\; \alpha = +1 , \vspace{2em} \\ 
  0 & \text{if}\; \alpha = -1.
  \end{cases}
  \label{eq:dilation-deriv}
\end{align}
Here, the derivative of $\jrc_\mathrm{m}$ with respect to $\sigma_{\cn}$ can be calculated as 
\begin{align}
  \dfrac{\pd \jrc_\mathrm{m}}{\pd \stress_{\cn}} = 
  \left \{ 
  \begin{array}{ll}
    \alpha \dfrac{\phi_{r}}{i_{\tau}^2} \dfrac{(\jrc_{p})_{\tau}}{\stress_{\cn} \ln 10} \dfrac{\pd \jrc_\mathrm{m}}{\pd (\phi_r / i_{\tau})}  & \text{if}\; 0 \leq \Lambda < \delta_{p}  , \vspace{1em}  \\ 
    0  & \text{if}\;   \delta_{p} \leq \Lambda ,    
  \end{array}
  \right.
\end{align}
where 
\begin{align}
  \dfrac{\pd \jrc_\mathrm{m}}{\pd (\phi_r / i_\tau)} = \dfrac{3(10 \Lambda - 3 \delta_{p})(\delta_{p} - 3 \Lambda)}{\left[3 \delta_{p} - (3 - 7 \phi_r / i_\tau)\Lambda \right]^2} . 
\end{align}
Lastly, the derivative of $M$ with respect to $\sigma_{\cn}$, which appears in Eq.~\eqref{eq:dilation-deriv}, can be calculated as 
\begin{align}
  \dfrac{\pd M}{\pd \stress_{\cn}} = \dfrac{\jrc_{p}}{12 \left[\log\left(\jcs/\stress_{\cn} \right) \right]^2} \dfrac{1}{\stress_{\cn} \ln 10} . 
\end{align}

\section*{Data Availability Statement} 
\label{sec:data-availability} 

The data that support the findings of this study are available from the corresponding author upon reasonable request.

\bibliography{references}

\end{document}